\documentclass[a4paper,fleqn,usenatbib]{mnras}
\usepackage{newtxtext,newtxmath}
\usepackage[T1]{fontenc}
\usepackage{ae,aecompl}

\usepackage{graphicx}	% Including figure files
\usepackage{amsmath}	% Advanced maths commands
\usepackage{amssymb}	% Extra maths symbols
\usepackage{pdfpages}   % include pdf 
\usepackage{acro}       % DeclareAcronym
\usepackage{cleveref}
\usepackage[normalem]{ulem}

\DeclareAcronym{eom}{
  short = EOM ,
  long = Equations of Motion ,
  short-plural = s,
}

\DeclareAcronym{smbh}{
  short = SMBH ,
  long = Supermassive Black Hole ,
  short-plural = s,
}
\DeclareAcronym{gw}{
  short = GW ,
  long = Gravitational Wave ,
  short-plural = ,
}

\DeclareAcronym{lisa}{
  short = {\it LISA} ,
  long = {\it Laser Interferometric  Space  Antenna},
  short-plural = ,
}

\DeclareAcronym{pn}{
  short = PN ,
  long = post-Newtonian ,
  short-plural = ,
}
\DeclareAcronym{bh}{
  short = BH ,
  long = Black Hole ,
  short-plural = s,
}

\DeclareAcronym{ligo}{
  short = LIGO ,
  long = Laser Interferometer Gravitational-Wave Observatory ,
  short-plural = ,
}
\DeclareAcronym{msp}{
  short = MSP ,
  long = millisecond pulsar ,
  short-plural = s,
}
\DeclareAcronym{emrb}{
  short = EMRB ,
  long = extreme-mass-ratio binary ,
  short-plural = s,
  long-plural-form = extreme-mass-ratio binaries,
}
\DeclareAcronym{imrb}{
  short = IMRB ,
  long = intermediate-mass-ratio binary ,
  short-plural = s,
  long-plural-form = intermediate-mass-ratio binaries,
}
\DeclareAcronym{emri}{
  short = EMRI ,
  long = extreme-mass-ratio-inspiral ,
  short-plural = s,
}
\DeclareAcronym{imri}{
  short = IMRI ,
  long = intermediate-mass-ratio-inspiral ,
  short-plural = s,
}
\DeclareAcronym{mpd}{
  short = MPD ,
  long = Mathisson-Papapetrou-Dixon ,
  short-plural = ,  
}

\DeclareAcronym{ska}{
  short = SKA,
  long = Square Kilometer Array,
}
\definecolor{grey}{gray}{0.35}

\title[Spin dynamics of a ms pulsar around a black hole]
{Spin dynamics of a millisecond pulsar orbiting closely around a massive black hole}

\author[K. J. Li et al.]{ 
Kaye Jiale Li,$^{1,2}$\thanks{E-mail: jlli@phys.cuhk.edu.hk (KJL), kinwah.wu@ucl.ac.uk (KW), 
 dinesh.singh@uregina.ca (DS) } 
Kinwah Wu,$^{1}$
and Dinesh Singh$^{3}$  \vspace*{0.25cm} \\ 
$^{1}$Mullard Space Science Laboratory, University College London, Holmbury St Mary, Surrey RH5 6NT, UK \\
$^{2}$Department of Physics, Chinese University of Hong Kong, Shatin, NT, Hong Kong SAR, China \\
$^{3}$Department of Physics, University of Regina, Regina, SK, S4S 0A2, Canada 
}

\date{Accepted 2019 February 5. Received 2019 February 5; in original form 2018 September 10}

\pubyear{2018}

\begin{document}
\label{firstpage}
\pagerange{\pageref{firstpage}--\pageref{lastpage}}
\maketitle

\begin{abstract}  
We investigate the spin dynamics of a \ac{msp} 
  in a tightly bounded orbit around a massive black hole.  
These { binaries} are progenitors of the 
  { \acp{emri} and \acp{imri}} 
  gravitational wave events. 
The \ac{mpd} formulation 
  is used to determine the orbital motion and spin modulation and evolution. 
We show that 
  the \ac{msp} will not be confined in a planar Keplerian orbit 
  and its spin will exhibit precession and nutation  
  induced by spin-orbit coupling and spin-curvature interaction. 
These spin and orbital behaviours will manifest observationally 
  in the temporal variations in the \ac{msp}'s pulsed emission 
  and, with certain geometries, in the self-occultation of the pulsar's emitting poles.  
Radio pulsar timing observations will be able to detect such signatures.   
These \acp{emrb} { and \acp{imrb}} are also strong gravitational wave sources. 
Combining radio pulsar timing and gravitational wave observations will  
  allow us to determine the dynamics of these 
  { systems} in high precision 
  and hence the subtle behaviours of spinning masses in strong gravity.   
\end{abstract}    


%%%%%%%%%%%%%%%%%%%%%%%%%%%%%%%%%%%%%%%%%%%%%%%%%%

%%%%%%%%%%%%%%%%% BODY OF PAPER %%%%%%%%%%%%%%%%%%

%%%%%%%%%%%%%%%%%%%%%%%%%%%%%%%%%%%%%%%%%%%%%%%%%%
%%%%%%%%%%%%%%%%%%%%%%%%%%%%%%%%%%%%%%%%%%%%%%%%%%
\section{Introduction}

The gravitational wave events, e.g. GW150914 \citep{Abbott2016} and 
  GW170608 \citep{Abbott2017b}, etc,  
  detected by \ac{ligo}
  provide strong support for Einstein's  theory of gravity, 
  i.e. general relativity (GR)  
  and evidence for astrophysical black holes. 
Although GR has passed a variety of tests in the weak field and strong field regimes, 
  there are still issues within it that require further clarification 
  \citep[see e.g.][]{Beiglbock1967,Costa2014}. 
Among them is the dynamics of spinning objects, in particular, 
  regarding how spin interacts with curved space-time 
  \citep[][]{Plyatsko1998,Iorio2012,Plyatsko2016} 
  and what the corresponding observable signatures are.     

%\ac{emrb} 
{ Binary} systems 
  containing an \ac{msp} orbiting around 
  a massive black hole (of $10^3 -10^6\;\!{\rm M}_\odot$) 
  are particularly useful for the study of spin-curvature interaction in GR.  
With the large mass ratio between the black hole and the \ac{msp},  
  the neutron star can be treated as a point test particle.   
The space-time is practically stationary, provided solely by the black hole.  
These allow us to construct models 
  that are simple enough to be mathematically tractable 
  yet sufficient for capturing the essences of the physics 
  and its subtle complexity. 
{ Depending on the mass of the black hole,
  the binary systems can be split explicitly
  into \acp{emrb} (for black holes between $10^5 - 10^6 \;\!{\rm M}_\odot$) 
  and \acp{imrb} (for black holes between $10^3 - 10^4 \;\!{\rm M}_\odot$),
  which correspond to different astrophysical systems.
\acp{emrb}/\acp{imrb} are progenitors of the \ac{emri}/\ac{imri} systems.
  They are major classes} of gravitational wave sources expected to be detected by 
  \ac{lisa} \citep[see e.g.][]{Amaro-Seoane2007}.  
The presence of an \ac{msp} guarantees 
  the electromagnetic counterparts of these { \acp{emrb}/\acp{imrb} 
  and the subsequent \ac{emri}/\ac{imri}} gravitational wave events. 
With high-precision radio timing observations 
  the spin and orbital dynamics of the \ac{msp} 
  can be investigated independently, 
  complimentary to the direct gravitational wave observations. 
  
\ac{emrb} { and \ac{imrb} systems are 
  astronomically important} in their own right.  
How { \acp{emrb}} were formed and how their progenitors had evolved to such configuration 
  are interesting questions to be answered. 
A possibility is that compact \ac{msp} - black hole binaries 
  were formed in very dense stellar environments \citep{Merritt2011,Clausen2014}, 
  e.g. the central region of 
  a { large stellar spheroid, 
  such as the core of a compact spheroidal galaxy,} 
  through sequences of stellar interactions.  
Another possibility is that they were produced 
  at the centre of a small elliptical or a Milky-Way-like spiral galaxy 
  when an \ac{msp} is captured by the nuclear black hole.  
{ \ac{imrb} systems 
  could also be formed in dense environments 
  where an intermediate-mass black hole capture an \ac{msp}. 
Globular clusters 
  are known to host a large population of pulsars, 
  in particular, \acp{msp} \citep[see e.g.][]{Lorimer2008}. 
Neutron stars are the more massive stars 
  in the globular clusters 
  and they would sink to the core of their host globular clusters 
  due to dynamical friction. 
If the globular cluster has 
  an intermediate-mass nuclear black hole,
  an \ac{imrb} system would, therefore, be formed.  
We will discuss the possibility of these events further in \S4.2.  
   }
  
Spinning neutron stars or spinning neutron-star binaries 
   revolving around a massive black hole 
   had been investigated in various astrophysical contexts 
   \citep[e.g.][]{Remmen2013,Singh2014,Rosa2015,Saxton2016}.   
Most of these studies put focus on the orbital dynamics 
  of the neutron star or the neutron-star binaries. 
This work will extend the previous investigations 
  to the dynamics of the neutron star's spin 
  when orbiting around a massive black hole 
  in the presence of spin-orbit and spin-curvature couplings. 
We determine on the observational signatures as diagnosis 
  and discuss their astrophysical and physical implications.  
The paper is organised as follows. 
In \S2 we present the formulation for the equation, 
  and in \S3 we show the results for systems 
  with parameters relevant to astrophysics 
  and to future pulsar timing observations 
  and gravitational wave experiment. 
Discussions on the astrophysics and physics implications 
  will be in \S4 and a summary in \S5.

%%%%%%%%%%%%%%%%%%%%%%%%%%%%%%%%%%%%%%%%%%%%%%%%%%
%%%%%%%%%%%%%%%%%%%%%%%%%%%%%%%%%%%%%%%%%%%%%%%%%%
\section{Equations of motion} 
\label{sec:EoM}

We adopt a $[\, - \;\! +\;\! +\;\! +\, ]$ signature for the metric 
 and a natural unit system, in which 
   the gravitational constant $G$ and the speed of light $c$ are unity ($G=c=1$). 
The \ac{msp}, a neutron star 
  with mass $m\;\!(=M_{\rm ns})$ and radius $R_{\rm ns}$,   
   orbits around a black hole of mass $M\;\!(=M_{\rm bh})$.  
The black hole has a Schwarzschild radius $R_{\rm sch} = 2M$, 
  and its rotation is specified by the spin parameter $a$,  
   with $a/M=1$ corresponding to a maximally rotating Kerr black hole 
   and $a/M=0$ corresponding to a non-rotating (Schwarzschild) black hole. 
The orbital separation between the \ac{msp} and the black hole, $r$, 
  is sufficiently large   
  such that $r > M \gg R_{\rm ns}>m$. 
The space-time is stationary, 
  determined by the black hole's gravity and rotation, i.e. a Kerr space-time.

The space-time interval, in the Boyer-Lindquist coordinates, is therefore given by  
\begin{equation} 
\begin{aligned} 
  - {\rm d}\tau^2 = & - \left(1- \frac{2Mr}{\Sigma}\right) {\rm d} t^2 -  
     \frac{4a M r \sin^2 \theta}{\Sigma}\;\! {\rm d} t \;\!{\rm d} \phi   
      \\  
    &  + \frac{\Sigma}{\Delta}\;\! {\rm d} r^2 + \Sigma\;\! {\rm d}\theta^2     
    +\left(r^2+a^2 +\frac{2a^2Mr \sin^2\theta}{\Sigma} \right)   \\
   & \times      \sin^2\theta \;\! {\rm d} \phi^2  \ ,   
\end{aligned} 
\label{eq-Kerr}  
\end{equation} 
   where $\Sigma = r^2 + a^2\cos^2 \theta$, 
   $\Delta = r^2 - 2Mr +a^2$ and 
   $(r,\theta, \phi)$ represents the spatial 3-vector 
   in the (pseudo-)spherical polar coordinates with the black-hole centre as the origin.   
The motion of the \ac{msp}, in the approximation as a particle-like object, 
  is determined by the continuity equation 
\begin{equation}
T^{\mu \nu}{}_{;\mu} = 0 \ ,
\end{equation}
   where the covariant derivative is taken with respect to the background spacetime.
For a spinning particle with 4-momentum $p^\mu$ and spin-tensor $s^{\mu \nu}$, 
  the continuity equation can be simplified to the
  \ac{mpd} equations:
\begin{equation}  
   \frac{{\rm D}p^\mu}{{\rm d}\tau}    =  -\frac{1}{2}\;\! {R^{\mu}}_{\nu\alpha \beta}u^\nu s^{\alpha \beta} \ ; 
 % + {\cal F}^{\mu}  \ ; 
\label{eq-MPD_a1}
\end{equation} 
\vspace*{-12pt}
\begin{equation} 
  \frac{{\rm D}s^{\mu \nu}}{{\rm d}\tau}   =  p^\mu u^\nu - p^\nu u^\mu   
  %+ {\cal T}^{\mu \nu} 
\label{eq-MPD_a2}
\end{equation}  
  \citep[see][]{Mashhoon2006,Plyatsko2011}, 
  where $u^\mu = {\rm d}x^\mu/{\rm d}\tau$ is the 4-velocity of the centre of mass. 
We have omitted the Dixon force ${\cal F}^\mu$  
  in the momentum evolutionary equation 
  and the Dixon torque ${\cal T}^{\mu \nu}$ 
  in the spin evolutionary equation 
  \citep[cf.][]{Singh2014}. 
They are arisen
  from the interaction of the quadrupole and higher-order mass moments 
  of the spinning object with the gravitational field  
  and therefore absent in the point-mass approximation 
  that we have adopted for the \ac{msp}.

To close the \ac{mpd} equation, a spin supplementary condition is required. 
We consider the Tulczyjew-Dixon (TD) condition  
   \citep[see][]{Tulczyjew1959,Deriglazov2017}, 
  where    
\begin{equation} 
  s^{\mu \nu} p_{\nu} = 0   \ . 
\end{equation}   
This, together with the point-mass approximation, 
  ensures that the mass of the \ac{msp}, given by     
\begin{equation}  
  m = \sqrt{-\;\!p^\mu p_\mu\;\!}   \  ,   
\end{equation} 
  is a constant of motion. 
The spin vector $s_\mu$ of the \ac{msp}
  is obtained by the contraction of the spin tensor $s^{\mu \nu}$:    
\begin{equation}  
     s_\mu = - \frac{1}{2m}\;\! \epsilon_{\mu \nu \alpha \beta}p^\nu s^{\alpha\beta} \ ; 
\label{eq-spin_1}
\end{equation} 
\vspace*{-12pt}
\begin{equation} 
   s^{\mu \nu } = \frac{1}{ m}\;\! \epsilon^{\mu \nu \alpha \beta} p_\alpha s_\beta  \ ,  
\label{eq-spin_2}  
\end{equation}  
  with Levi-Civita tensor 
   $\epsilon_{\mu \nu \alpha \beta} = \sqrt{-\;\! g\;\!}\;\! \sigma_{\mu \nu \alpha \beta}$  
   adopting the $\sigma_{0123} = +1$ permutation. 
Contraction of the spin vector gives the scalar 
\begin{equation} \label{eq-SpinScalar} 
  s^2 = s^\mu s_\mu = \frac{1}{2}\;\! s^{\mu \nu}s_{\mu \nu } \ , 
\end{equation}  
  which is a constant of motion.   
  
%%%%%%%%%%%%%%%%%%%%%%%%%%%%%%%%%%%%%%%%%%%%%%%%%% 
% Figure 1
\begin{figure}
%	  \vspace*{0.05cm}   \center 
%          \vspace*{6cm}          
    \centering
    \vspace*{-0.5cm}
    \includegraphics[width=1\columnwidth]{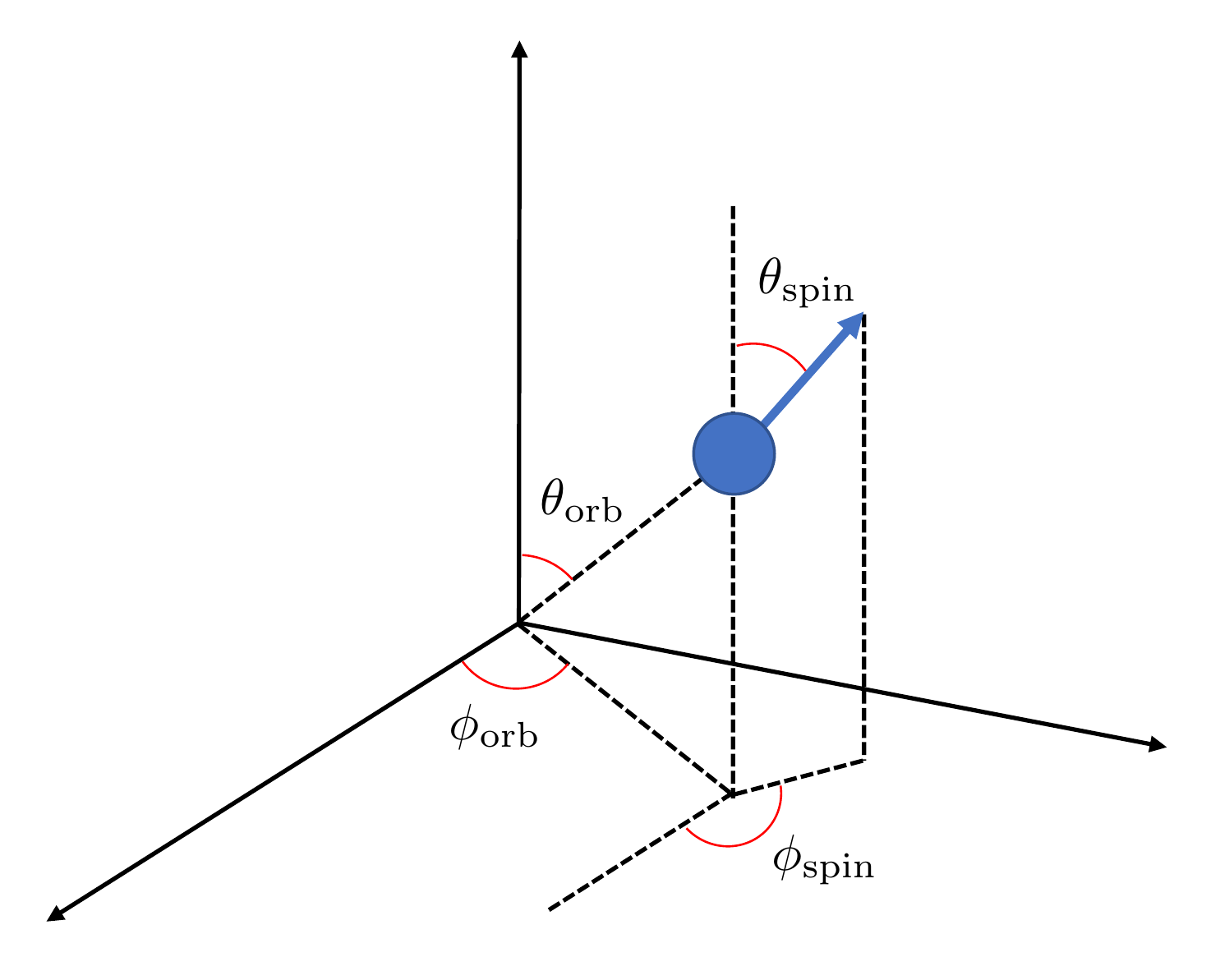}
    \vspace*{-0.5cm}
    \caption{The geometry of the system. The orbital angles $\phi_{\rm orb}$ and $\theta_{\rm orb}$ are defined with respect to the (pseudo-)spherical polar coordinate 
      with respect to the centre of the black hole. 
     The spin angles $\phi_{\rm spin}$ and $\theta_{\rm spin}$ are defined on a Cartesian coordinate in the \ac{msp}'s local tetrad frame.}
    \label{fig:geometry}
\end{figure}
%%%%%%%%%%%%%%%%%%%%%%%%%%%%%%%%%%%%%%%%%%%%%%%%%%

In the regime where the M{\o}ller radius of the \ac{msp} (i.e., a neutron star)     
\begin{equation} 
  r_{\rm M} = \frac{s}{m} \ll r \ ,  
\end{equation} 
  the dipole-dipole interaction and the higher-order multipole interactions,
  which are much weaker than the pole-dipole interaction, can be ignored. 
Thus,    
\begin{equation} 
  \left( \frac{p^\mu}{m} - u^\mu \right) \sim \frac{M \;\!  {{r_{\rm M}}^2}}{r^3}  \ll 1 \ ,   
\end{equation} 
 and the approximation scheme proposed by \cite{Chicone2005} is applicable. 
With $p^\mu \approx m u^\mu$, 
  the \ac{mpd} equations are reduced to    
\begin{equation} 
  \frac{{\rm D}u^\mu}{{\rm d}\tau} = -\frac{1}{2m}\;\! {R^\mu}_{\nu \alpha\beta}u^\nu s^{\alpha \beta} \ ;  
\label{eq-MPS_b1}
\end{equation} 
\vspace*{-10pt}
\begin{equation} 
    \frac{{\rm D}s^{\mu \nu}}{{\rm d}\tau} \approx 0  \ ,  
\label{eq-MPD_b2}
\end{equation}  
 and the closure condition becomes  
\begin{equation} \label{eq-closure}
  s^{\mu \nu} u_\nu \approx 0   
\end{equation} 
\citep{Chicone2005,Mashhoon2006}.  
This is essentially the Frenkel-Mathisson-Pirani (FMP) condition \citep{Frenkel1926,Mathisson1937,Costa2014,Costa2018} 
\footnote{{ It has been pointed out that, the motion of a particle under different
  spin supplementary conditions are equivalent to dipole order \citep{Costa2014}.
Such equivalences were not shown for the evolution of spin.   
We look forward to future work about this issue.}}. 

To investigate the difference 
  between the cases with and without consideration of spin-curvature coupling,  
  we introduce a parameter switch $\lambda$ into the \ac{mpd} equations  
  as in \cite{Singh2014} \citep[see also][]{Singh2005}:     
\begin{equation}  
\frac{{\rm d} p^\alpha}{{\rm d}\tau} = - {\Gamma^\alpha}_{\mu \nu} p^\mu u^\nu 
   + \lambda \left( \frac{1}{2m} {R^\alpha}_{\beta \rho \sigma} {\epsilon^{\beta\sigma}}_{\mu \nu} s^\mu p^\nu u^\beta   \right) \ ;  
\label{eq:mpd-momentum}
\end{equation} 
\vspace*{-10pt}
\begin{equation} 
\frac{{\rm d} s^\alpha}{{\rm d}\tau} = - {\Gamma^\alpha}_{\mu \nu} s^\mu u^\nu  
 +  \lambda \left( \frac{1}{2m^3} R_{\gamma \beta \rho \sigma} {\epsilon^{\beta\sigma}}_{\mu \nu} 
    s^\mu p^\nu s^\gamma u^\beta   \right) p^\alpha \ ; 
\label{eq:mpd-spin}
\end{equation} 
\vspace*{-10pt}
\begin{equation}  
\frac{{\rm d} x^\alpha}{{\rm d}\tau} = u^\alpha =  - \frac{p^\delta u_\delta}{m^2} \left( p^\alpha 
  + \frac{1}{2} \; \! \frac{\lambda(s^{\alpha\beta} R_{\beta\gamma\mu\nu}p^\gamma s^{\mu \nu})}{m^2 
     +\lambda(R_{\mu \nu \rho \sigma} s^{\mu \nu}s^{\beta \sigma}/4)  } \right)    \ . 
\label{eq:orb-coup}     
\end{equation}   
Spin-curvature coupling is included when $\lambda =1$, and excluded when $\lambda =0$.   
In this formula, the spin 4-vector is Fermi-Walker transported along the worldline of the centre-of-motion of the \ac{msp}.

%%%%%%%%%%%%%%%%%%%%%%%%%%%%%%%%%%%%%%%%%%%%%%%%%%
%%%%%%%%%%%%%%%%%%%%%%%%%%%%%%%%%%%%%%%%%%%%%%%%%%
\section{Spin and orbit modulation of the millisecond pulsar} 
\label{sec:modulation}

We adopt a neutron-star mass $m =1.5~{\rm M}_\odot$.  
The \ac{msp} spin period $P_{\rm s}$ is 
   taken to be $1~{\rm ms}$
(with spin $s = 0.3787 m^2 $ 
\footnote{ The spin angular momentum of { the} \ac{msp} depends on the 
  internal structure of the \ac{msp}, which is model-dependent. 
{ Here we assume that the \ac{msp} is a uniform
  solid sphere with radius $ R_{\rm ns} = 10\;\!{\rm km}$. 
Under such an approximation, the \ac{msp} has a spin $s = \frac{2}{5} m R_{\rm ns}^2 \frac{2 \pi}{ P_{\rm s}} \approx 0.3787 m^2$. }  } 
throught out this paper)
, for the extremely fast rotating \ac{msp}  
\footnote{A comprehensive catalogue of pulsars 
   in Galactic globular clusters complied by P.~Freire can be found in www.naic.edu/~pfreire/GCpsr.html.}  
   \citep[see][for the period distributions of \ac{msp}]{Papitto2014,Ozel2016}.  
The massive black hole (MBH) is taken to have mass $M = 10^3$, $10^4$ and $10^5\;\!{\rm M}_\odot$,   
  and the spin parameter $a/M = 0$, $\pm0.5$ and $\pm 0.99$.   
The orbit of the \ac{msp} around the black hole is bounded, 
  with $``+"$ and $``-"$ signs in $a/M$ corresponding  
  to  the black hole in 
  a prograde and a retrograde rotation with respect to the orbital motion of the \ac{msp}.  
The semi-major axis $r$ of the orbit of \ac{msp}, 
  defined as the mean of minimum and maximum distances between the \ac{msp} and black hole, is chosen to be 20, 50 and $100\;\!M$. 
The eccentricity $e$ of the orbit is calculated 
  using the method described in Appendix~\ref{EccentricityCalculation}. 
It has values ranging between 0 and 0.6 
\footnote{{ It worth noticing that, depending on the formation channels,
  some \acp{emri}/\acp{imri} may possess zero eccentricity \citep[see e.g.][]{Miller2005}.
Other mechanisms, for example, compact stars driven by gravitational radiation 
  (i.e. gravitational bremsstrahlung) \citep{Quinlan1989} or 
  stars on orbits near the loss cone \citep{Hopman2005}
  could possess large }{ eccentricities. 
The evolution of such highly eccentric \acp{emrb} and \acp{imrb} are driven by gravitational radiation,
  and the interaction with other stars can be ignored \citep{Konstantinidis2013}. 
These studies presented distribution of initial orbital eccentricity of \acp{emrb} and \acp{imrb}
  when they enter the \ac{lisa} bandwidth 
  \citep[i.e. when their orbital periods are about $10^4 \;\!{\rm sec}$, 
  as defined in][hereafter ``initial eccentricity'' and ``initial semi-major axis'' refer to this criteria]{Hopman2005}.
When they enter the relativistic regime that we are interested in, the orbits are greatly circularised 
  by the emission of \ac{gw}. 
For example, using the two-body radiation formula in \citep{Peters1964},     
  for a \ac{imri} with $10^3 \;\!{\rm M}_\odot$, and initial semi-major axis $2.25 \times 10^{-7} {\rm pc}$,
  eccentricity $0.998$ \citep[adapted from Fig.~5 of][notice that this is not necessarily a reliable result, as pointed out in their paper]{Hopman2005}, 
  the eccentricity is reduced to $\sim 0.6$ when semi-major axis is reduced to $20 M$ ($M = 10^3 \;\!{\rm M}_\odot$).
%In general, for $10^3, 10^4$ and $ 10^5 \;\!{\rm M}_\odot$ central black holes, the eccentricity of the compact stars
%  orbiting around the black hole will be smaller than $0.4$ if their initial eccentricities  
%  are smaller than $0.996$, $0.98$ and $0.93$, respectively,
In general, for $10^3, 10^4$ and $ 10^5 \;\!{\rm M}_\odot$ central black holes, the eccentricity of the compact stars
  orbiting around the black hole will be smaller than $0.6$ if their initial eccentricities  
  are smaller than $0.998$, $0.99$ and $0.96$, respectively.
Therefore, we would like to restrict the eccentricity to be between $0$ and $0.6$. 
}}.
The initial orientation of the \ac{msp}'s spin axis 
  is set to be $0^\circ$, $45^\circ$ and $90^\circ$ 
  with respect to the initial Newtonian orbital angular momentum.
The system geometry is shown in Fig.~\ref{fig:geometry}.  

%%%%%%%%%%%%%%%%%%%%%%%%%%%%%%%%%%%%%%%%%%%%%%%%%% 
% Figure 2
\begin{figure}
	  \vspace*{0.05cm}   \center 
      \includegraphics[width=1\columnwidth]{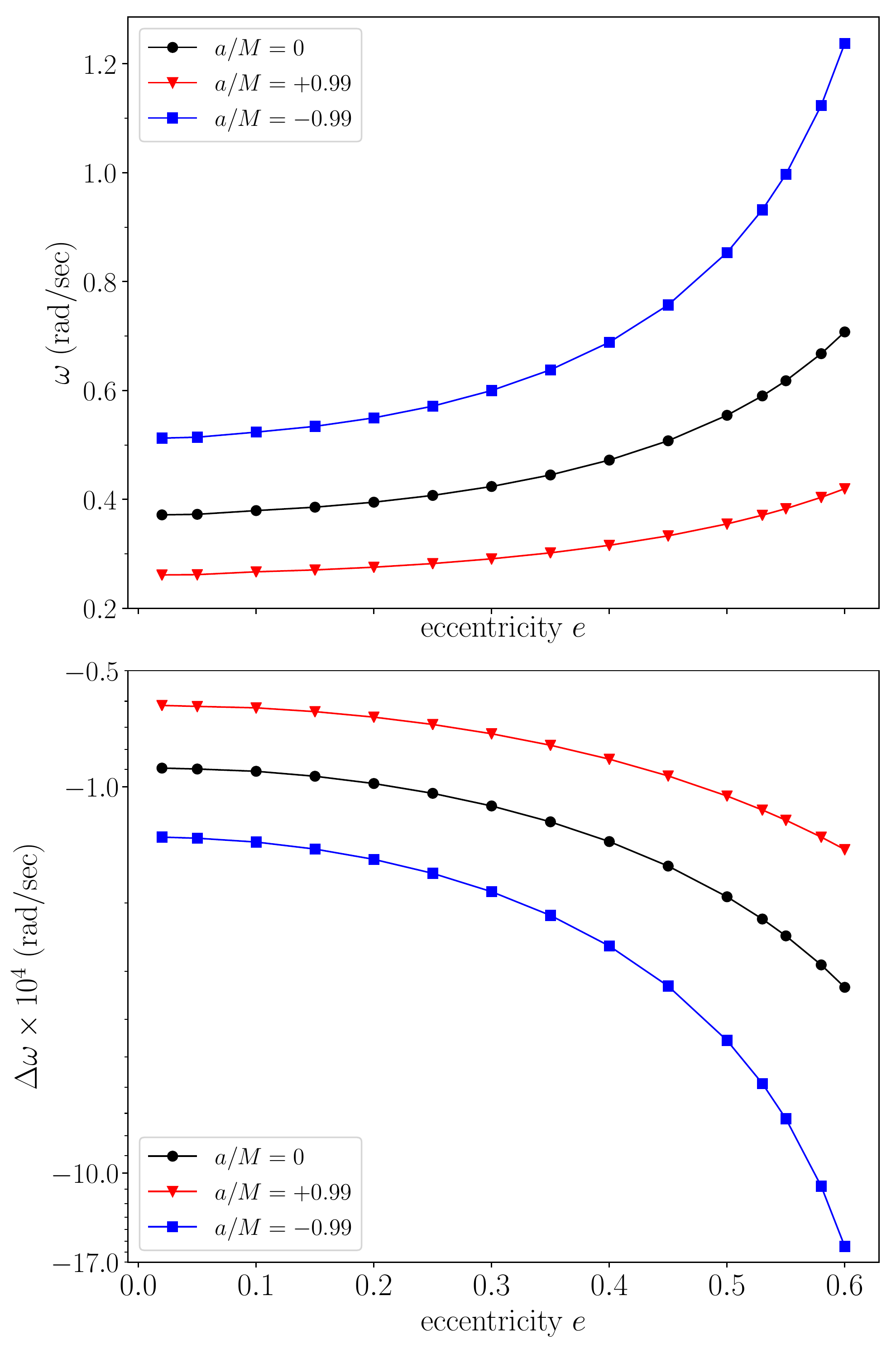}
        \vspace*{-0.25cm}          
    \caption{{ (Figure is updated)}
% the parameters of each figures
The upper panel shows the rate of periastron advance as a function of eccentricity $e$ in the pure geodesic case. 
{ The lower panel shows the corrections to the rate of periastron advance 
  due to spin-orbit and spin-curvature couplings,
  for an \ac{msp} with period $1~{\rm ms}$.}  
The semi-major axis of the orbit of the \ac{msp} is set to be $r=20 M$, 
 and the mass of the central black hole $10^3\;\!{\rm M}_{\odot}$. 
The blue, black, red lines correspond to the cases of the black-hole spin
   $a/M=-0.99$ (retrograde with respect to the \ac{msp} orbit), $0$, and $0.99$ (prograde with respect to the \ac{msp} orbit) respectively.
The spin of the \ac{msp} is parallel to the orbital angular momentum initially. 
% compare with different bh masses 
The rate of geodetic periastron advance in the upper panel is decreased by $10$ and $10^2$ times 
  for central black holes with $10^4\;\!{\rm M}_{\odot}$ and 
  $10^5\;\!{\rm M}_{\odot}$ respectively, 
  while the correction to the rate due to the \ac{msp}'s spin is decreased by $10^2$ and $10^4$ times respectively.  
{ Notice that the y-axis of the lower panel is in log scale,
  which shows that the effects of spin-orbit and spin-curvature couplings are greatly enhanced in the highly eccentric cases.}
}
    \label{fig:Periastron}
\end{figure}
%%%%%%%%%%%%%%%%%%%%%%%%%%%%%%%%%%%%%%%%%%%%%%%%%%

%%%%%%%%%%%%%%%%%%%%%%%%%%%%%%%%%%%%%%%%%%%%%%%%%%
\subsection{Results}  
\label{sec:results}
% 1st part, mainly talk about the orbital precession

In the \ac{mpd} formulation, the \ac{msp}'s orbital motion 
  and spin evolution are interdependent (see Eq.~\ref{eq:orb-coup}). 
Fig.~\ref{fig:Periastron} shows the rate of the periastron advancement  
  as a function of the orbital eccentricity ($e$) expected 
  for geodesic motion (top panel) 
  and the correction to the rate when spin-orbit and spin-curvature couplings are considered (bottom panel).  
The precession rate of the \ac{msp}'s orbit is generally faster for $a/M<0$ than for $a/M>0$.    
The precession is determined by several mechanisms, 
  among them the strongest is due to geodesic motion,    
  similar to that in Mercury when it revolves around the Sun.   
Another one is due to the Lense-Thirring effect, 
  arisen from the black hole's rotation.  
This effect is clearly visible 
  when comparing the rates for non-zero $a/M$ with that for $a/M =0$. 
The precession is also contributed  
  by the interaction between the \ac{msp}'s spin 
  with the \ac{msp}'s orbit motion 
  and with the space-time curvature induced by the black hole's gravity.  
Note that for the system parameters considered in this work,  
  the advancement of the orbital precession is comparable (see top panel, Fig.~\ref{fig:Periastron}) 
  to the angular velocity of the \ac{msp}'s orbital motion, 
  which is about $\sim 2~{\rm rad~s}^{-1}$ (for $M = 10^3 \;\!{\rm M}_{\odot}$). 
Perturbation methods, in particular those assuming a quasi-circular orbit, 
  are therefore not always applicable 
  when determining the \ac{msp}'s orbital dynamics 
  for systems with non-zero eccentricities. 

In { a} classical eccentric binary system,   
  orbital precession is usually caused by tidal interactions between the components 
  and/or the presence of quadrupole and/or higher-order multipole mass moments 
  in the components.   
{ 
Here we have demonstrated the presence of the well-known 
  orbital precession due to general relativistic effects,
  such effects have been studied extensively in literature \citep[see e.g.][]{Kidder1995,Ruangsri2016}.    
}
One of the consequence is that the orbital precession is enhanced. 
This additional acceleration can be illustrated 
  in terms of a 1PN (first-order post-Newtonian) correction 
  for a parametrised Keplerian binary system 
  \citep{Damour1985,Damour1986}.  
  
The strength of spin-orbit interaction in the system 
  may be characterised in terms of an effective interaction 
\begin{equation} \label{EffectiveSpin}
      \chi_{\rm eff} = \frac{1}{(M +m) } \left( \frac{\boldsymbol{s}}{m} 
      + \frac{\boldsymbol{S}_{\rm MBH}}{M}  \right) \cdot \boldsymbol{\hat{L}} \ , 
\end{equation}
   where $\boldsymbol{s}$ and $\boldsymbol{S}_{\rm MBH}$ are the spin vectors of the \ac{msp} and the massive black hole respectively 
   and $\boldsymbol{\hat L}$ is the unit directional vector of the \ac{msp}'s orbital angular momentum $\boldsymbol{L}$.  
Here, $\boldsymbol{S}_{\rm MBH}$ is related to the spin parameter by  
  $\boldsymbol{S}_{\rm MBH} = a M \boldsymbol{\hat{S}}_{\rm MBH}$.  
(Hereafter, unless otherwise stated, $\boldsymbol{\hat x}$ denotes that the unit directional vector of a vector $\boldsymbol{x}$.) 
For a black hole and an \ac{msp} that have the same value of dimensionless spin $a/M$ and $s/m^2$, 
  the effects of \ac{msp}'s spin on the orbital dynamics and spin dynamics are scaled with the factor $m/M$. 
The value of dimensionless spin of \ac{msp} depends on its rotational period and inner structure.
From the observational perspective, 
  the pulsed signals allow us to determine the rotational period of the \ac{msp}, 
  while the strength of spin-orbit couplings depends on the dimensionless spin of the \ac{msp}.   
Therefore, by measuring such a binary system, we can not only probe the space time structure of the MBH, but also achieve two independent measurements of the \ac{msp}'s rotation period and moment of inertia, which potentially provide us clues about the inner structure of the \ac{msp}.

\begin{figure}
	  \vspace*{0.05cm}   \center 
     \includegraphics[width=1\columnwidth]{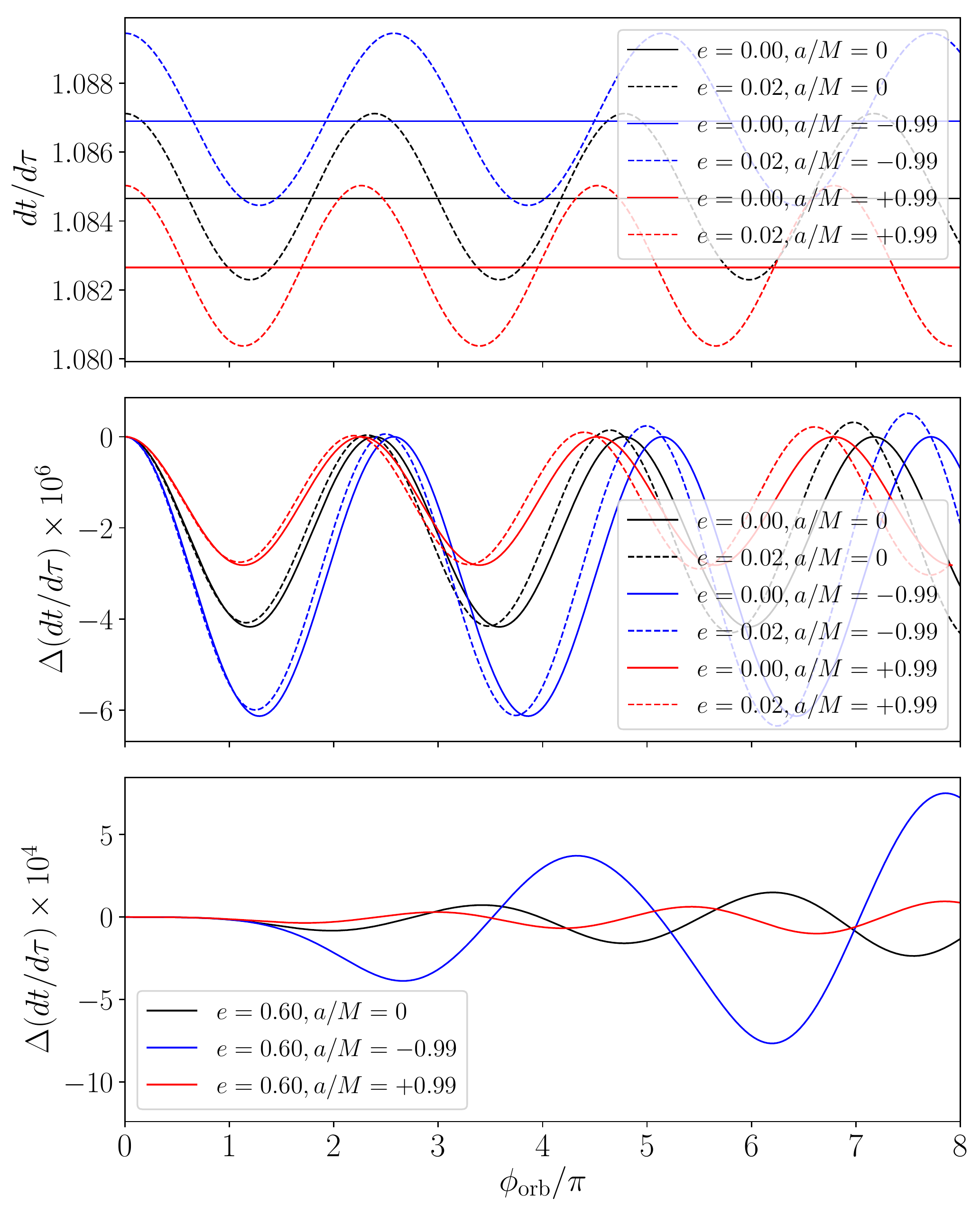}
        \vspace*{0cm}          
\caption{{ (Figure is updated)}
    The time-like component of \ac{msp}'s 4-velocity $u^0 = {\rm d}t/{\rm d}\tau$ of 4-velocity of the \ac{msp}, 
    and the corrections due to \ac{msp}'s spin-orbit and spin-curvature coupling. 
 Upper panel shows the geodesics ${\rm d}t/{\rm d}\tau$ of circular and quasi-circular orbit. Middle panel is the corrections due to spin-orbit and spin-curvature coupling for circular and quasi-circular orbits. Lower panel demonstrate the corrections in ${\rm d}t/{\rm d}\tau$ due to spin-orbit coupling for elliptical orbits. Note that, the scale of $y$-axis are different for all three panels. The initial spin is $\theta_{\rm spin}=\pi/4$, leaning in the direction towards black hole. All other parameters are the same as in Fig.~\ref{fig:Periastron}.
}
    \label{fig:ProperTime}
\end{figure}

Fig.~\ref{fig:ProperTime} shows the corresponding difference of $u^0$, 
 the ratio of the coordinate time ${{\rm d}t}$ and the proper time ${{\rm d}\tau}$ of the \ac{msp}. 
The \ac{msp} serves as an accurate clock \footnote{The gravitation effects on clocks associated with a spinning object  
  in a circular orbit around a gravitating mass were studied by 
  \cite{Bini2005}.  }. 
Therefore, ${{\rm d}t}/{{\rm d}\tau}$ can be directly measured 
  if we know the intrinsic rotational period of the \ac{msp}. 

We consider a quasi-circular orbit approximation  
  and determine the different effects on $u^0$ 
  by expanding the analytic formula of $u^0$ for geodesic orbits 
  with respect to the eccentricity $e$
  and the Post-Newtonian (PN) factor $M/r$.
As no assumption is made for the spins of the black hole and the \ac{msp}, 
  the expression that we obtain is valid for the extremely rotating black hole 
   (with $a/M = \pm 0.99$) and fast spinning \ac{msp}. 

The details of the calculations are shown in Appendix~\ref{ProperTimeCalculation}.  
The estimated values of $\Delta u^0 $ due to each factor are shown in Table.~\ref{table:Scale}. 
These results are consistent  
  with the those shown in the upper and middle panels of Fig.~\ref{fig:ProperTime}. 

For an \ac{msp} with a highly eccentric orbit, 
   only the corrections due to the coupling of \ac{msp}'s spin to orbital angular momentum are shown in Fig.~\ref{fig:ProperTime} (lower panel). 
The corrections are much greater than the results estimated by using the linear approximation in Appendix~\ref{ProperTimeCalculation},
  mainly due to the breaking down 
  of Taylor expansion of $u^0$ with respect to the eccentricity $e$. 
In general, a correction of about $\sim 20\;\!{\rm ns}$ arises within a duration of $\sim 10-1000 ~{\rm sec}$,
  for black hole with masses between $10^{3} - 10^{5} \;\!{\rm M}_\odot$.   
% This effect is noticeable for future pulsar timing observations..

%%%%%%%%%%%%%%%%%%%%%%%%%%%%%%%%%%%%%%%%%%%%%%%%%%
% Table 1 
\begin{center}
\begin{table}
\begin{tabular}{ |p{1.7cm}|p{2cm}|p{1.5cm}|p{1.5cm}|} 
 \hline
 Effect & scale & $\Delta  \frac{{\rm d}t}{{\rm d}\tau}$ & ${\Delta  \frac{{\rm d}t}{{\rm d}\tau}}/ {T_{\rm orb}} $\\
  \hline 
    \hline 
GR & $\frac{3}{2} \frac{M}{r}$  & $0.075~{\rm ms}$ & $2.7 \times 10^{-5}$ \\
\hline 
Eccentricity & $ 2 e  \frac{M}{r} $ & $2~\mu{\rm s}$ & $7.2 \times 10^{-7} $ \\
\hline
MBH's spin & $3 \frac{a}{M} (\frac{M}{r} )^{5/2} $ & $1.66~\mu{\rm s}$ & $6.0 \times 10^{-7} $\\
\hline 
\ac{msp}'s spin & $ 2 \delta e \frac{M}{r}  + \frac{3}{2}  \frac{M}{r}   \frac{\delta r}{r}  $ & $ 3.12~{\rm ns}$  & $1.1 \times 10^{-9}$\\
 \hline
\end{tabular}
 \caption{
The correction to the time component of \ac{msp}'s 4-velocity (i.e. $u^0$), which is the ratio of coordinate time ${{\rm d}t}$ and proper time ${{\rm d}\tau}$, by different factors for circular and quasi-circular orbits. The formula are calculated in Appendix~\ref{ProperTimeCalculation}, and only leading orders are shown in the table for order estimation. Here we take $m =1.5~{\rm M}_\odot$, $M=10^3~{\rm M}_\odot$, semi-major axis $r=20M$, $e=0.02$, $a/M= \pm 0.99$ and $s=0.4 m^2$ for an example. 
For the effect of eccentricity, ${\rm d}t/{\rm d}\tau$ is evaluated at either periastron or apastron, and is compared to circular orbit with the same semi-major axis.
For the effect of MBH's spin, ${\rm d}t/{\rm d}\tau$ is evaluated for circular orbit around spinning MBH, and compared to circular orbit with the same radius around non-spinning MBH. 
When we include the \ac{msp}'s spin, the eccentricity is perturbed by $\delta e \sim 2 \times 10^{-5}$, while the semi-major axis is perturbed by $\delta r = 3 \times 10^{-4} M$, leading to a perturbation of about $3.12~{\rm ns}$. The method used in the table is only valid for circular and quasi-circular orbits, and the estimated order is consistent with upper and middle panels of Fig.~\ref{fig:ProperTime}. The effect of \ac{msp}'s spin would be greatly underestimated for highly eccentric orbits using the method here, compared with the exact numerical results of \ac{mpd} equations, as shown in lower panel of Fig.~\ref{fig:ProperTime}. 
 } \label{table:Scale}
 \end{table}
\end{center}
%%%%%%%%%%%%%%%%%%%%%%%%%%%%%%%%%%%%%%%%%%%%%%%%%%

%%%%%%%%%%%%%%%%%%%%%%%%%%%%%%%%%%%%%%%%%%%%%%%%%%
% Figure 3
%%%%%%%%%%%%%%%%%%%%%%%%%%%%%%%%%%%%%%%%%%%%%%%%%%

% spin precession Fig 4 and Fig 5
When the axis of the spin is not aligned with the angular momentum, the spinning axis, 
  as well as the orbital angular momentum undergoes precession.
This effect was investigated previously, either based on
  a one-graviton interaction analog \citep[e.g.][]{Barker1966,Barker1979}, or by 
  approaches similar to this work \citep[e.g.][]{Bini2005}. 
As the orbital angular momentum, as well as the spin, precesses around the total angular momentum, 
  when such in-plane spin is present, 
  the orbital angular momentum is no longer constant, in either magnitude or direction. 
The precession of the orbital plane, as a consequence, is usually referred to as out-of-plane motion \citep{Singh2014}.
 
The focus of this work is on the spin dynamics, and we will present the numerical results and the observables in the following. 
Typically, there are three different motions related to the spin: 
  precession, nutation and rotation,  
  corresponding to three Euler angles respectively.
The choice of Euler angles is subject to the choice of the observer and reference frame. 
Here we choose the first Euler angle (which describes the precession) to be $\phi_{\rm spin}$,
  and the second Euler angle (which describes the nutation) to be $\theta_{\rm spin}$.
The third Euler angle would be of interest for modelling the motion of the magnetic field axis with respect to the spin axis. 
As the time scale of rotation of magnetic field axis is $P_{\rm s} = 1~{\rm ms}$, 
  which is much smaller than the time scale of precession and nutation, 
  we will not include its effect until Sec.~\ref{sec:observation}. 

\begin{figure}
	  \vspace*{0.05cm}   \center 
     \includegraphics[width=1\columnwidth]{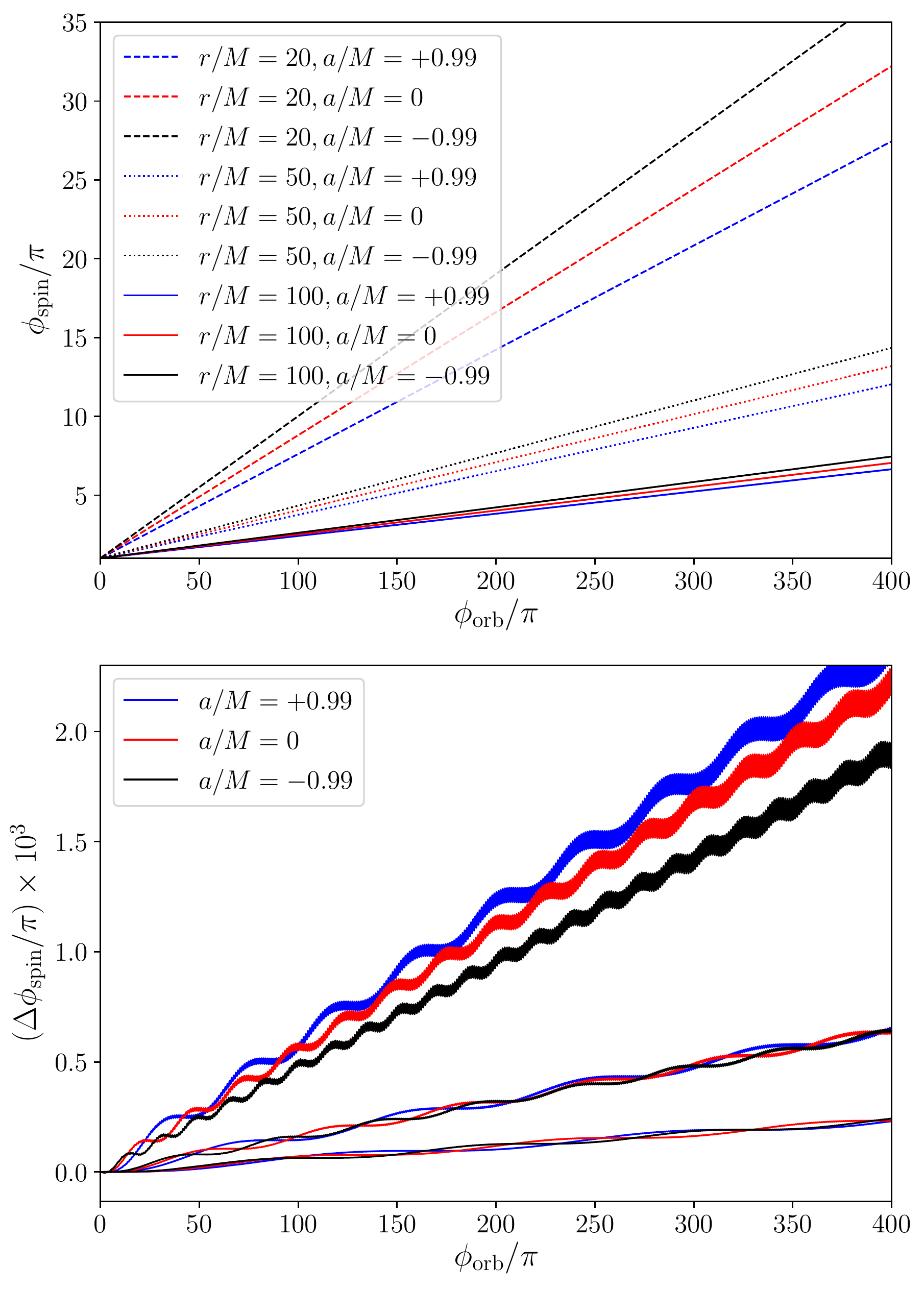}
        \vspace*{0cm}          
    \caption{
% the parameters of each figures
The upper panel shows the precession in $\phi_{\rm spin}$ 
{ for circular and quasi-circular geodetic orbits}.
The { semi-major axis of the orbit of the \ac{msp} is} $r=20\;\!M$ (dashed lines), 
  $r=50\;\!M$ (dotted line) and $r=100\;\!M$ (solid line). 
The black, red, blue lines correspond to the black hole spin $a/M=-0.99$ (i.e. retrograde \ac{msp} orbit), $0$, and $0.99$ (i.e. prograde \ac{msp} orbit). 
The eccentricity of the \ac{msp} orbit is approximately  $0$, 
  and the central black hole is of mass $10^3\;\!M_{\odot}$.
The initial spin of the neutron star is inclined at an angle $\theta_{\rm spin} = \pi/4$,
  and leaning in the direction toward the black hole (i.e. $\phi_{\rm spin} = \pi$).  
The other parameters are the same as those in Fig.~\ref{fig:Periastron}, 
The lower panel shows the correction for the precession $\phi_{\rm spin}$ in the presence of spin-orbit coupling, 
  { and the orbital parameters are the same as the corresponding ones in the upper panels}.    
}
    \label{fig:SpinPrecessionPhi}
\end{figure}

The precession of the spinning axis of \ac{msp} is shown in the upper panel of Fig.~\ref{fig:SpinPrecessionPhi}. 
Despite that, in this figure, the mass of the massive black hole is $10^3\;\!{\rm M}_{\odot}$, 
  the upper panel is valid for BHs with larger masses. 
The reason is that, on the Newtonian order the spin $ \boldsymbol{s}$ of the \ac{msp} evolves as 
\begin{equation}
\begin{aligned} \label{eq-PrecessionOfSpin}
\boldsymbol{\dot{s}} =  \frac{1}{r^3}\left( \frac{3M}{2m}
    (\boldsymbol{L} \times \boldsymbol{s}) 
- \boldsymbol{S_{\rm MBH}} \times \boldsymbol{s} + 3(\boldsymbol{\hat{n}} \cdot \boldsymbol{S_{\rm MBH}})( \boldsymbol{\hat{n}} \times  \boldsymbol{s})  \right)  
\end{aligned}
\end{equation}  
   \citep[see][]{Barker1979,Thorne1985,Kidder1995}, 
    where $\boldsymbol{L}$ is the orbital angular momentum, 
    the leading order of which is the Newtonian orbital angular momentum 
    $\boldsymbol{L_{\rm N}} = m \boldsymbol{r} \times \boldsymbol{v}$. 
The precession frequency due to the Newtonian angular momentum is 
\begin{equation}
\begin{aligned} \label{eq:spinprecessionnewtonian} 
\omega_{L_{\rm N}} = \frac{1}{r^3}  \frac{3M}{2m}\;\! 
 \left|\boldsymbol{L_{\rm N}}\right| \propto \frac{M }{r^2} v \ .
\end{aligned}
\end{equation}
The orbital frequency is however $\omega_{\rm orb} \propto {v}/{r}$, 
  which differs from the above precession frequency by a factor of ${M}/{r}$. 
Therefore, the ratios of spin's precession velocities and orbital velocities remain the same for different black hole masses $M$ with the same $r/M$. 
As shown in the upper panel of Fig.~\ref{fig:SpinPrecessionPhi}, 
  the precession rates descend, $\propto {M}/{r}$, with increasing radius $r$ 
  for all the cases. 

The differences in the precession velocity for the spinning black holes with respect to that of the Schwarzschild black hole, 
  as shown in the upper panel of Fig.~\ref{fig:SpinPrecessionPhi}  
  are due to the second term of Eq.~\ref{eq-PrecessionOfSpin}. 
The precession frequency caused by $\boldsymbol{S}_{\rm MBH}$ is 
\begin{equation}
\begin{aligned} 
\omega_{S_{\rm MBH}} = \frac{1}{r^3}  
 \left|\boldsymbol{S_{\rm MBH}}\right| 
 \propto \frac{a}{M} \frac{M^2}{r^3} \propto \frac{a}{M} \sqrt{\frac{M}{r}} \omega_{L_{\rm N}} \ .
\end{aligned}
\end{equation}
The relation is consistent with that resulted from the \ac{mpd} equation, 
  which is shown in the upper panel of Fig.~\ref{fig:SpinPrecessionPhi}, 
  despite that the derivations of the two are not based on identical assumptions. 

The lower panels of Fig.~\ref{fig:SpinPrecessionPhi} demonstrate the combined effect of the spin-orbit and spin-spin couplings  
  on the spin precession rate of the \ac{msp}. 
The effect is non-linear and is not easily seen from the Eq.~\ref{eq-PrecessionOfSpin}.
However, we can estimate the order of magnitude of it on the spin precession rate in terms of an effective spin (Eq.~\ref{EffectiveSpin}), which may be expressed as 
\begin{equation}
\begin{aligned} \label{eq:approximatespinprecession}
\omega_{s} \sim \frac{1}{r^3}  \frac{M}{m} \left| \boldsymbol{s} \right|  
\propto \frac{s}{m^2} \frac{m}{M} \frac{M^2}{r^3} \propto \frac{s}{m^2} \frac{m}{M} \omega_{L_{\rm N}} \ . 
\end{aligned}
\end{equation}
As there is an ${m}/{M}$ dependence, 
  this spin coupling cannot be ignored especially for systems that are consist of intermediate-mass-ratio binaries.

%%%%%%%%%%%%%%%%%%%%%%%%%%%%%%%%%%%%%%%%%%%%%%%%%%
% Figure 4
%%%%%%%%%%%%%%%%%%%%%%%%%%%%%%%%%%%%%%%%%%%%%%%%%%

\begin{figure}
	  \vspace*{0.05cm}   \center 
     \includegraphics[width=1\columnwidth]{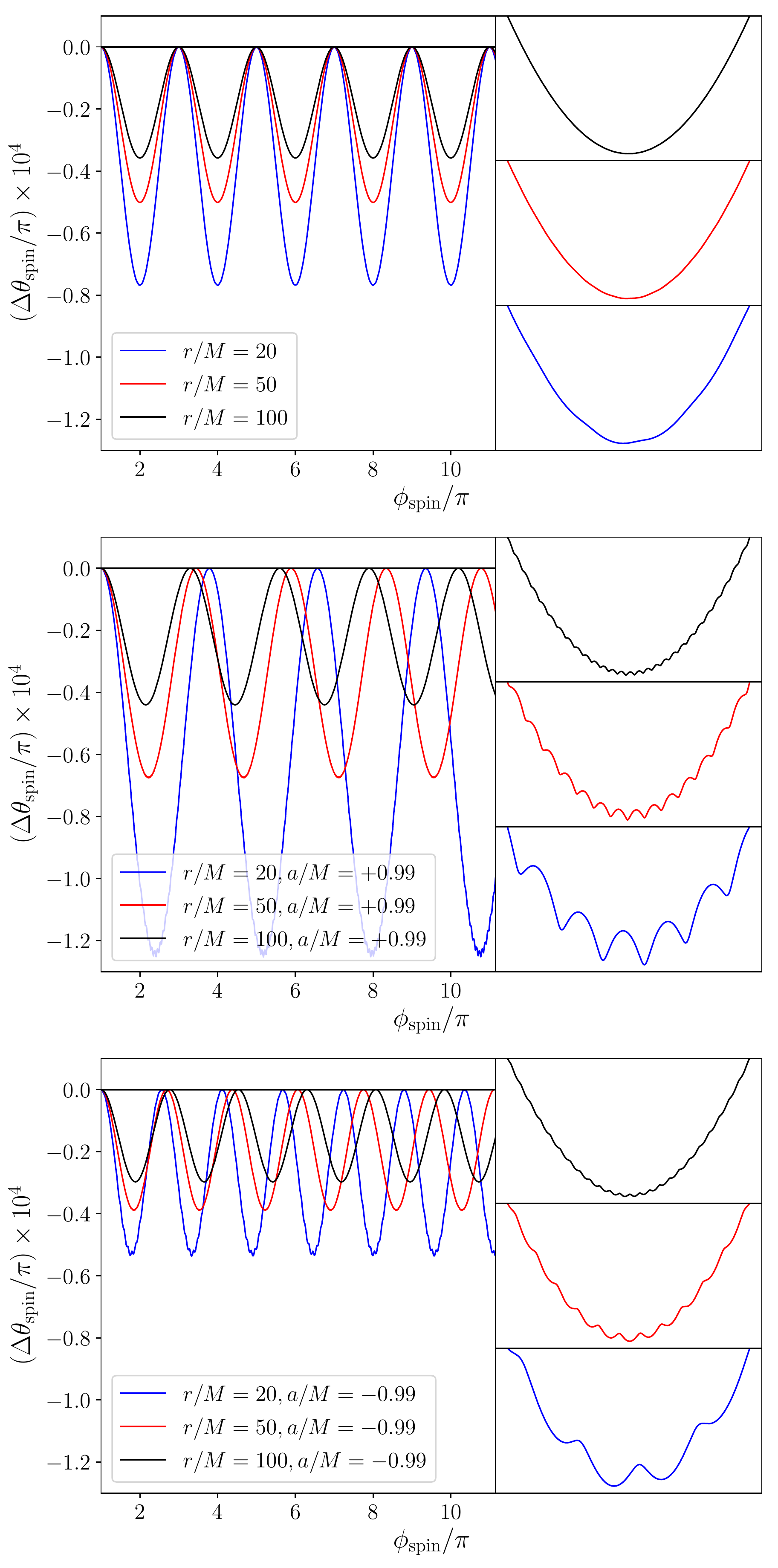}
        \vspace*{0cm}          
    \caption{
% the parameters of each figures
The nutation of $\theta_{\rm spin}$ for the cases with $a/M = 0$ 
  (upper panel), 
  $0.99$ (left side, middle panel) and $-0.99$ (lower panel),  
  corresponding respectively to the red, blue and black curves in Fig.~\ref{fig:SpinPrecessionPhi} respectively.  
The small amplitude nutations contributed by the spin-spin interaction 
  are shown correspondingly on the right side of each panel. 
In this figure, the cases with $r=20\;\!M$, 
  $r=50\;\!M$ and $r=100\;\!M$ are respresnted by blue, red and blue respectively. 
The initial spin of the \ac{msp} is inclined at an angle $\theta_{\rm spin} = \pi/2$ and leaning in the direction toward the black hole (i.e. $\phi_{\rm spin} = \pi$). 
The other parameters are the same as those in Fig.~\ref{fig:SpinPrecessionPhi}. 
The horizontal straight line at $\theta_{\rm spin}=0$ in each panel is the reference of no nutation, 
 the geodesic case. 
%The small figures on the right-hand-side of each panel are rescaled plots of the valleys of the lines of the same color 
%  in the corresponding left-hand-side panel. 
 }
    \label{fig:SpinPrecessionTheta}
\end{figure}

The nutation of the \ac{msp}'s spinning axis  
  is caused by the combination of the geodesics effect (which only involve the space-time around the black hole), 
  and the coupling of \ac{msp}'s spin to the orbital angular momentum and to the MBH's spin. 
The nutation due to the geodesic effect and the corrections due to \ac{msp}'s spin are shown in Fig.~\ref{fig:SpinPrecessionTheta}.
For Schwarzschild black holes (upper panel, Fig.~\ref{fig:SpinPrecessionTheta}), 
   the total angular momentum 
\begin{equation} \label{eq-AngularMomentum}
\boldsymbol{J} =  \boldsymbol{L} + \boldsymbol{s} = \boldsymbol{L}_{\rm N} + \boldsymbol{L}_{\rm PN} + \boldsymbol{L}_{\rm SO} + \boldsymbol{L}_{\rm 2PN} 
 + \boldsymbol{s} \ 
\end{equation} 
   \citep{Kidder1995} is conserved at $2$PN order. 
When the angular momentum $\boldsymbol{L}_{\rm N}$ wobbles,
  the orbital plane of the \ac{msp} will tilt accordingly, 
  and the \ac{msp} spin axis will also wobbles around, 
  changing the direction and magnitude of the spin 3-vector ${\boldsymbol s}$ of the \ac{msp}.  
 
The \ac{msp} orbital angular momentum is an external angular momentum.  
It is conserved when there is a rotational symmetry. 
This symmetry is however broken in the presence of the \ac{msp} spin. 
The situation is slightly different for the \ac{msp} spin, 
  as it is an intrinsic feature of the \ac{msp}.  
  
The change of the magnitude of \ac{msp} spin's 3-vector is a unique phenomenon,  
  revealed in the \ac{mpd} equations, 
  whilst it would remain constant in the usual PN formulations \citep[]{Barker1979,Thorne1985,Kidder1995}. 
Although these two descriptions of the evolution of spin seem to be contradictory, 
  they are, in fact, consistent with each other.
In the PN formulations, the evolution equation of a particle's spin is usually written as 
  an outer product of a external angular momentum vector with the particle's spin vector (e.g. Eq.~\ref{eq-PrecessionOfSpin}), 
  which directly implies the conservation of the spin's magnitude.
This, however, is written in the comoving frame of the particle,
  and it has been shown to be equivalent to Fermi-Walker transport equation \cite[see][]{PastorLambare2017}.
By contrast, the \ac{mpd} formulation is written in the distant observer's frame \footnote{Sometimes
the \ac{mpd} formula is converted into comoving frame to avoid the ambiguity 
  in the definitions of the time component of the \ac{msp} spin's 4-vector \citep[see][]{Damour2008}.}. 
It seems that the different choices of reference frame can account for 
  the difference between \ac{mpd} formulation and PN formulations. 
But there are still ambiguity in the definition of spin 4-vector, 
  and the physical meaning of the time component $s^0$ is not well understood. 
By using the analogy between spin and relativistic angular momentum, 
  it can be shown that $s^0$ is related to the dynamic mass moment \footnote{Dynamic mass moment 
  is defined as $\boldsymbol{N} = m \boldsymbol{x} - t \boldsymbol{p}$, \citep[see e.g.][]{Penrose2004}.}, 
  i.e. the offset of centre of mass and centre of momentum, measured by the comoving observer.

Recall the closure condition Eq.~\ref{eq-closure}, which can also be written as $p^\alpha s_\alpha = 0$ \citep{Costa2014} and hence
\begin{equation}
u^\alpha s_\alpha = 0 \ .
\end{equation}
Dividing both sides by $u^0$, the 4-velocity is converted into velocity with respect to the coordinate time:
\begin{equation} \label{eq-TimeComponentOfSpin}
s_0 = -\left( s_1 \frac{{\rm d} r}{{\rm d}t} + s_2 \frac{{\rm d} \theta}{{\rm d}t} + s_3 \frac{{\rm d} \phi}{{\rm d}t} \right) \ ,
\end{equation}
where $s_{\mu}$ are the components of dual vector, defined as $s_{\mu} = g_{\mu \nu} s^{\nu}$. 
Eq.~\ref{eq-TimeComponentOfSpin} describes the projection of spin 3-vector onto the velocity measured by a local static observer. 
Besides, the factor $u^0$ describes the time dilation effect, therefore, $s_0$ is also related to the relativistic light aberration as described in \cite{Rafikov2006}.

\begin{figure}
	  \vspace*{0.05cm}   \center 
     \includegraphics[width=1\columnwidth]{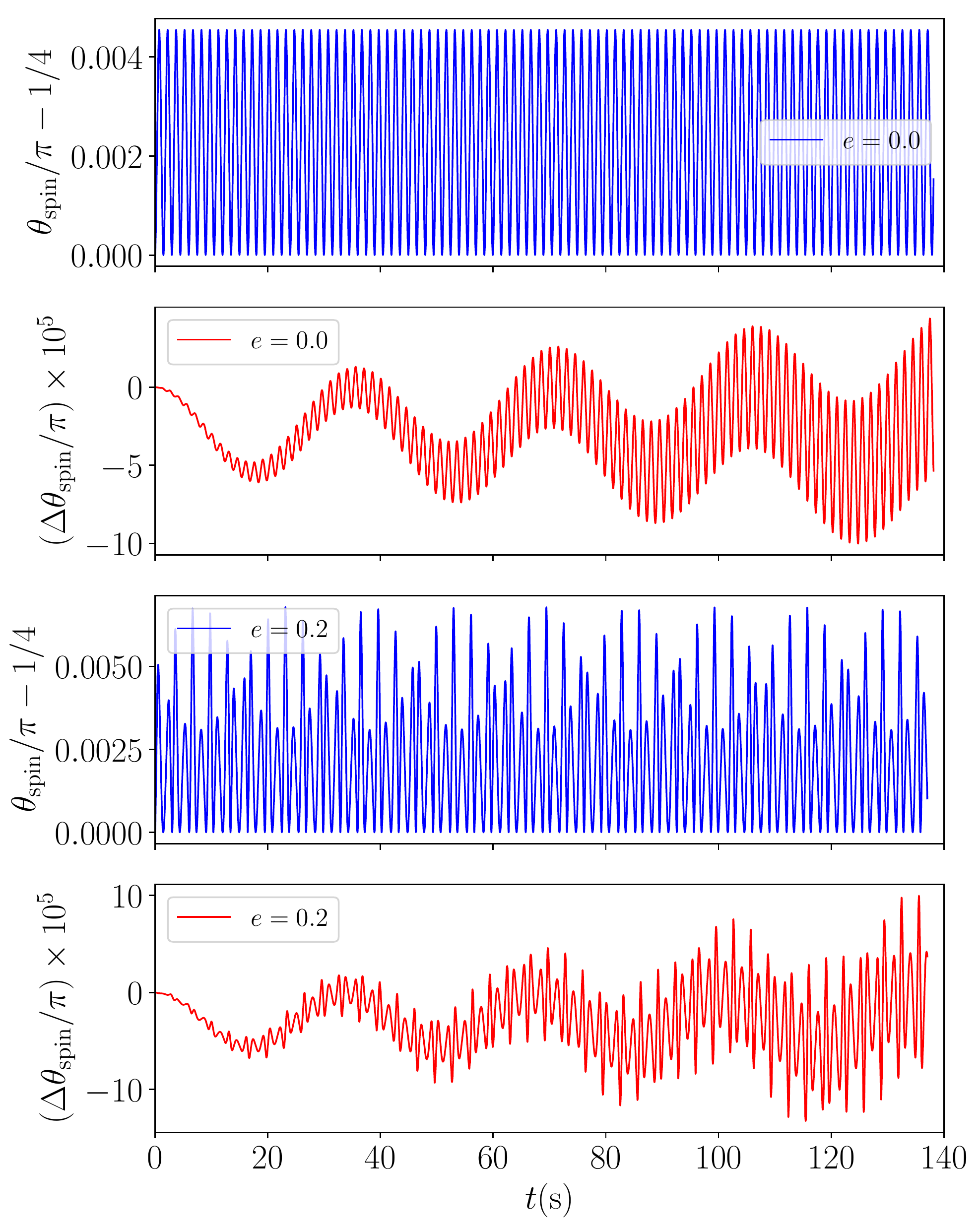}
        \vspace*{-0.25cm}          
    \caption{
Nutation of spin axis due to geodesics (blues lines, 1st and 3rd panels from above), 
  and the corrections to nutation due to the coupling of \ac{msp}'s spin. 
In the upper two panels, the eccentricity of the \ac{msp}'s orbit is $e=0$, 
  in the lower two panels, $e=0.2$. 
The MBH's spin is $a/M=0$, and the semi-major axis of the orbit of \ac{msp} are $r=20\;\!M$. 
Other parameters are the same as those used in Fig.~\ref{fig:SpinPrecessionPhi}
}
    \label{fig:SpinPrecessionTheta2}
\end{figure}

Fig.~\ref{fig:SpinPrecessionTheta} and Fig.~\ref{fig:SpinPrecessionTheta2} show 
  the nutation of spin axis of the \ac{msp} due to the spin-orbit coupling. 
In Fig.~\ref{fig:SpinPrecessionTheta}, 
we selected the special cases $\theta_{\rm spin} = \pi/2$, 
  where \ac{msp}'s spin is within the orbital plane, 
  for an example. 
When $\lambda = 0$, the spin rotates around the orbital angular momentum 
  and the spinning angular momentum of the MBH, 
  therefore, there is only change in the first Euler angle $\phi_{\rm spin}$. 
This explains why in all three panels, when $\lambda=0$, there is no nutation. 
Nevertheless, when we include spin-orbit coupling, a small nutation occurs. 
Such an nutation cannot be explained by Eq.~\ref{eq-PrecessionOfSpin}, 
  nor the first order correction to it by replacing $\boldsymbol{L}$ wth $\boldsymbol{L_{\rm N}} + \boldsymbol{L_{\rm PN}}$, 
  as both of them would lead to vanishing projection of $\dot{{\boldsymbol s}}$ onto the $\boldsymbol{L_{\rm N}}$ direction. 
Thus we need higher order corrections, which, naturally, 
  explains why the order of nutation due to the spin-orbit coupling 
  is smaller than the precession due to spin-orbit coupling by an order. 

In fact, most of the orbital components of total angular momentum in Eq.~\ref{eq-AngularMomentum} 
  are parallel to $\boldsymbol{L_{\rm N}}$, and the only one that could account for such nutation is $\boldsymbol{L_{\rm SO}}$. 
From \cite{Kidder1995}, we have 
\begin{equation}
\begin{aligned} 
\label{eq-SOAngularMomentum}
\boldsymbol{L_{\rm SO}} = & \frac{m}{M} \bigg\{ 
\frac{M}{r} \boldsymbol{\hat{n}} \times \bigg[ \boldsymbol{\hat{n}} \times \left( 3 (\boldsymbol{s} + \boldsymbol{S}_{\rm MBH}) + (m-M) \left(\frac{\boldsymbol{S_{\rm MBH}}}{M}-\frac{\boldsymbol{s}}{m}\right) \right)
\bigg] \\
& - \frac{1}{2} \boldsymbol{v} \times \bigg[ \boldsymbol{v} \times \left(\boldsymbol{s} + \boldsymbol{S_{\rm MBH}} + (m-M) \left(\frac{\boldsymbol{S_{\rm MBH}}}{M}-\frac{\boldsymbol{s}}{m}\right) \right) \bigg] 
\bigg\} \ .
\end{aligned}
\end{equation}
The scale of it is:
\begin{equation}
\begin{aligned}
\left|\boldsymbol{L_{\rm SO}}\right| \propto & 
\frac{s}{m^2} \frac{m^2 M}{r} \simeq \frac{s}{m^2}  \frac{m}{M} \sqrt{\left(\frac{M}{r}\right)^3}\ 
 \left|\boldsymbol{L_{\rm N}}\right| \ . 
\end{aligned}
\end{equation}
Therefore the frequency of nutation due to $\boldsymbol{L_{\rm SO}}$ is 
\begin{equation}
\begin{aligned} 
\omega_{L_{\rm SO}} =  \frac{1}{r^3}  \frac{3M}{2m} |\boldsymbol{L_{\rm SO}}| \propto  \frac{s}{m^2}  \frac{m}{M} \frac{M^3}{r^4}   \propto \frac{s}{m^2}  \frac{m}{M} \sqrt{\left(\frac{M}{r}\right)^3}\  \omega_{L_{\rm N}} \ .
\end{aligned}
\end{equation}

In Eq.~\ref{eq-SOAngularMomentum}, when we set $\boldsymbol{S_{\rm MBH}}$ to be zero, the only part that contributes to the nutation is
\begin{equation}
\begin{aligned} \label{eq-SOAngularMomentum2}
\boldsymbol{L}_{\rm SO} \simeq & \frac{m}{M} \bigg\{ 
\frac{M}{r}  \frac{M}{m} \boldsymbol{\hat{n}} \times \left[ \boldsymbol{\hat{n}} \times \boldsymbol{s} \right]  - \frac{1}{2} \frac{M}{m} \boldsymbol{v} \times \left[ \boldsymbol{v} \times \boldsymbol{s} \right]
\bigg\} \ , \\
\simeq & \frac{M}{r} \bigg\{  \boldsymbol{\hat{n}} 
\left[ \boldsymbol{\hat{n}} \cdot \boldsymbol{s} \right] - \frac{1}{2}  \boldsymbol{\hat{v}} \left[ \boldsymbol{\hat{v}} \cdot \boldsymbol{s} \right] \bigg\} + \text{terms that are parallel to } \boldsymbol{s} \ .
\end{aligned}
\end{equation}
The time dependencies of $\boldsymbol{\hat{n}} ( \boldsymbol{\hat{n}} \cdot  \boldsymbol{s})$ and $\boldsymbol{\hat{v}} ( \boldsymbol{\hat{v}} \cdot  \boldsymbol{s})$ are the same for circular orbits. 
Take $\boldsymbol{\hat{n}} ( \boldsymbol{\hat{n}} \cdot  \boldsymbol{s})$ for an example. 
Assuming the orbital angular frequency to be $\omega$, the angular frequency of the spin's precession to be $\Omega$, we have $\Omega \ll \omega$. The value of $\boldsymbol{\hat{n}} ( \boldsymbol{\hat{n}} \cdot  \boldsymbol{s})$ averaged over an orbital period is 
\begin{equation}
\begin{aligned} \label{eq-nutationfrequency}
\langle \boldsymbol{\hat{n}} ( \boldsymbol{\hat{n}} \cdot \boldsymbol{\hat{s}}) \boldsymbol{\hat{x}} \rangle = 
  & \int  \frac{{\rm d}t}{T}   \cos (t \omega ) \big[\sin (t \omega ) \sin (t \Omega  )+\cos (t \omega ) \cos (t \Omega )\big] \\
\sim & \frac{\sin (t (2 \omega -\Omega ))}{2 (2 \omega -\Omega )}+\frac{\sin (t \Omega )}{2 \Omega } + \text{constants} \ . 
\end{aligned}
\end{equation} 
The nutation therefore has two frequencies, 
  with the dominate one having the same frequency as the precession of the spin axis  
  (as shown in the upper panel of Fig.~\ref{fig:SpinPrecessionTheta}). 
The other frequency is not  apparent in the Fig.~\ref{fig:SpinPrecessionTheta}, 
  as the precession velocity $\Omega$ varies with frequency $2(\omega - \Omega)$, 
   which cancels out the first term of Eq.~\ref{eq-nutationfrequency}. 

We have not considered the black hole spin explicitly in the above discussion. 
When we include the MBH's spin, the situation is much more complicated. 
Here we present only the numerical results  
  for the case with an initial spin orientation $\theta_{\rm spin}=\pi/2$, 
   in  Fig.~\ref{fig:SpinPrecessionTheta}.  
There are also small amplitude nutation resulting from spin-spin corrections, 
  besides the corrections to the period of spin's nutation with respect to spin's precession, 
  and they are shown in the small figures in the right side of each panel. 

%%%%%%%%%%%%%%%%%%%%%%%%%%%%%%%%%%%%%%%%%%%%%%%%%%
% Figure 5
%%%%%%%%%%%%%%%%%%%%%%%%%%%%%%%%%%%%%%%%%%%%%%%%%%

The nutation due to spin-orbit coupling 
  for a general case, with $\theta_{\rm spin}=\pi/4$, 
  is shown in Fig.~\ref{fig:SpinPrecessionTheta2}. 
It demonstrates that the spinning axis of the \ac{msp} undergoes nutation 
  even without spin-orbit coupling. 
This nutation comes from Thomas precession 
  (or equivalently Eq.\ref{eq-PrecessionOfSpin} in the distant observer's frame). 
As shown in Fig.~\ref{fig:SpinPrecessionTheta2} (upper panel), 
  the geodesic nutation is affected by the alignment of the \ac{msp}'s velocity 
  and its spinning axis in the distant observer's frame,  
  and it has a angular frequency roughly of $2(\omega - \Omega)$. 
The amplitude is approximated by a Lorentz transformation from the comoving frame to the local static frame, and has value
\begin{equation}
\begin{aligned}
\Delta \theta_{\rm spin} \simeq \tan ^{-1}\left(\frac{2 (\gamma -1) \sin ^2\left(\theta_{\rm spin} \right)}{(\gamma -1) \sin \left(2 \theta_{\rm spin}\right)+2}\right) \ ,
\end{aligned}
\end{equation}
    where $\gamma = ( 1-{M}/{r})^{-1/2}$ is the Lorentz factor, $\theta_{\rm spin}$ is the initial angle of spin. 
As shown in the upper panel of Fig.~\ref{fig:SpinPrecessionTheta2}, for $\theta_{\rm spin} = \pi/4$ and $r/M = 20$, 
   $\Delta \theta_{\rm spin} \simeq 0.004\;\! \pi$, 
   a value that is consistent with that  obtained by solving the \ac{mpd} equation directly. 
When the spin-couplings are included in the \ac{mpd} equation (i.e. $\lambda =1$ in Eq.~\ref{eq:mpd-momentum},\ref{eq:mpd-spin},\ref{eq:orb-coup}),
the corrections have two frequencies. 
The lower frequency is similar to the one shown in Fig.~\ref{fig:SpinPrecessionTheta} and Eq.~\ref{eq-nutationfrequency}, 
  while the higher frequency is due to the shift of the precession velocity, 
  and therefore roughly has a frequency of $2\;\!(\omega - \Omega)$. 
The modulations of the $\Delta \theta_{\rm spin}$ amplitude 
  shown in the lower two panels of Fig.~\ref{fig:SpinPrecessionTheta2}   
  are the consequences of the variations of $\gamma$ over the orbital cycle. 

%%%%%%%%%%%%%%%%%%%%%%%%%%%%%%%%%%%%%%%%%%%%%%%%%%
% Figure 6
%%%%%%%%%%%%%%%%%%%%%%%%%%%%%%%%%%%%%%%%%%%%%%%%%% 

%%%%%%%%%%%%%%%%%%%%%%%%%%%%%%%%%%%%%%%%%%%%%%%%%%
%%%%%%%%%%%%%%%%%%%%%%%%%%%%%%%%%%%%%%%%%%%%%%%%%%
\section{Discussion}   
\label{sec:discussion}  

%%%%%%%%%%%%%%%%%%%%%%%%%%%%%%%%%%%%%%%%%%%%%%%%%%
%%%%%%%%%%%%%%%%%%%%%%%%%%%%%%%%%%%%%%%%%%%%%%%%%%
\subsection{Observational prospects} 
\label{sec:observation}

The results presented have several observational prospects. 
% What is the observables due to the corrections ?
% eccentricity
The waveform of gravitational wave emitted by a eccentric binary system has been a heated topic 
  in both equal and extreme mass ratio systems \citep{Favata2014,Kavanagh2017,Moore2018}. 
\acp{emri}{ / \acp{imri}} are expected to possess large orbital eccentricities when
  they enter the \ac{lisa} frequency band \citep[see][]{Amaro-Seoane2007,Amaro-Seoane2018}.
Ignoring the orbital eccentricity may lead to systematic biases 
  in parameter estimation of compact binary gravitational wave sources \citep{Favata2014} 
  and also to loss in the source detection \citep{Moore2018}. 
\ac{lisa} is expected to be more sensitive to the orbital eccentricity of the binary systems 
  than \ac{ligo}, 
  and the problem would therefore be severe.  
Despite the technical difficulties, modelling the waveforms 
 of the eccentric binary systems is essential, 
  as \acp{emri}{ / \acp{imri}} with high eccentricity are favourable target systems of \ac{lisa} 
   -- the high eccentricity leads to stronger signal and hence an enhancement of detectable events 
      \citep{Barack2004,Amaro-Seoane2015}.
Constructing waveform templates with high accuracy is therefore a very crucial objective 
  in the preparation of future \ac{lisa} observations, 
  as well as in full exploitation of the \ac{ligo} capability.

% more about the eccentricity...
In this calculation, energy dissipation is not  considered,  
  and hence its effect on the eccentricity evolution is not included. 
The time scale of eccentricity evolution is however comparable to the time scale of orbital decay 
  \citep[see][]{Peters1964}. 
It therefore justifies our approximation that the eccentricity does not vary on the time scale of spin precession and nutation. 
% back to quasi-circular cases
% Besides the effects related to eccentricity, 
Note that the spin-orbit coupling will introduce a shift in the phase of the gravitational wave emitted, and  
  the accumulative effect of spin on the phase of gravitational wave of the circular system 
{ was studied by \citep{Burko2015,Warburton2017,Fujita2018}. 
}

\begin{figure}
	  \vspace*{-0.05cm}   \center 
	  \includegraphics[width=1\columnwidth]{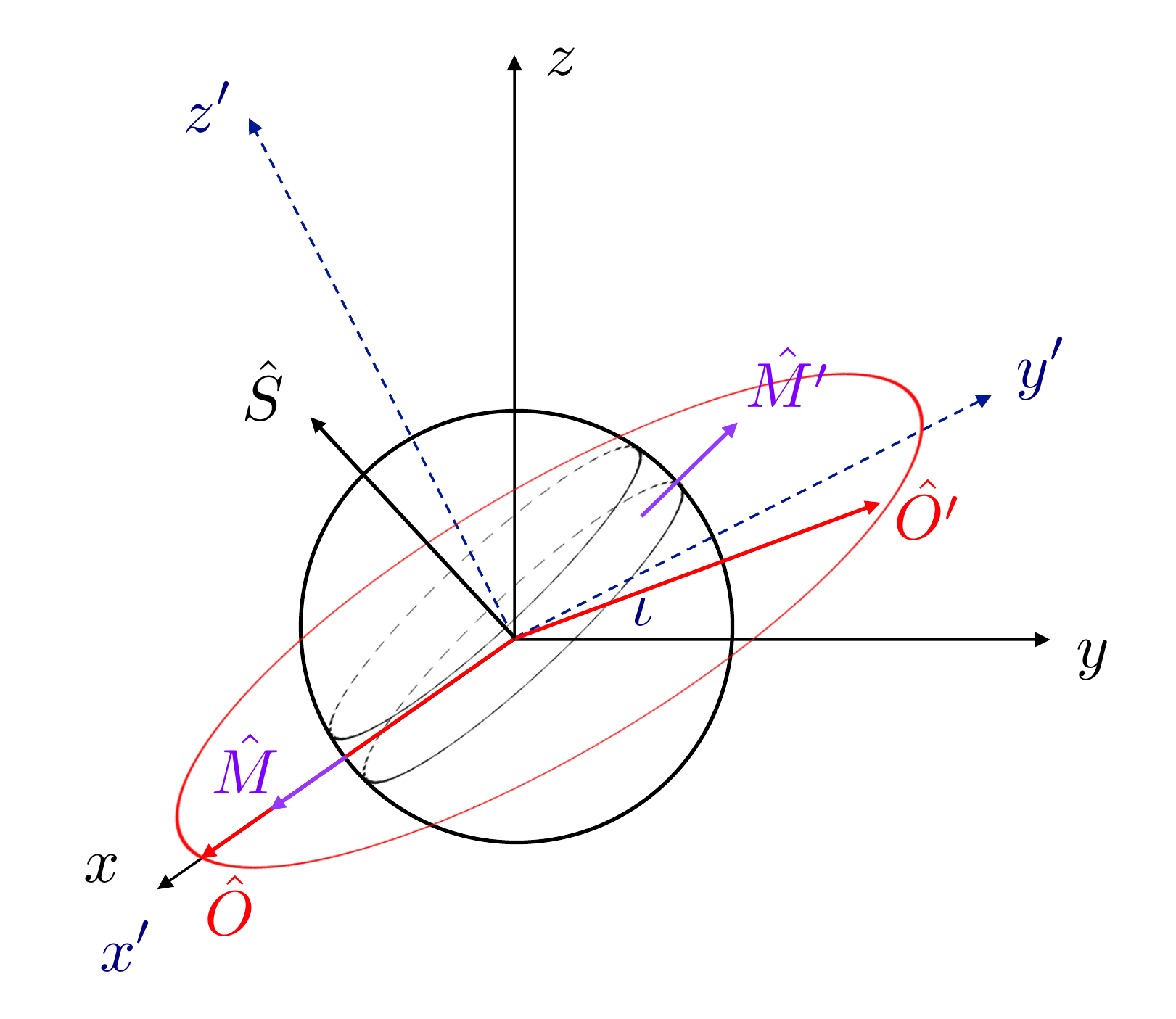}
           %\vspace*{-1.05cm}          
	  \includegraphics[width=1\columnwidth]{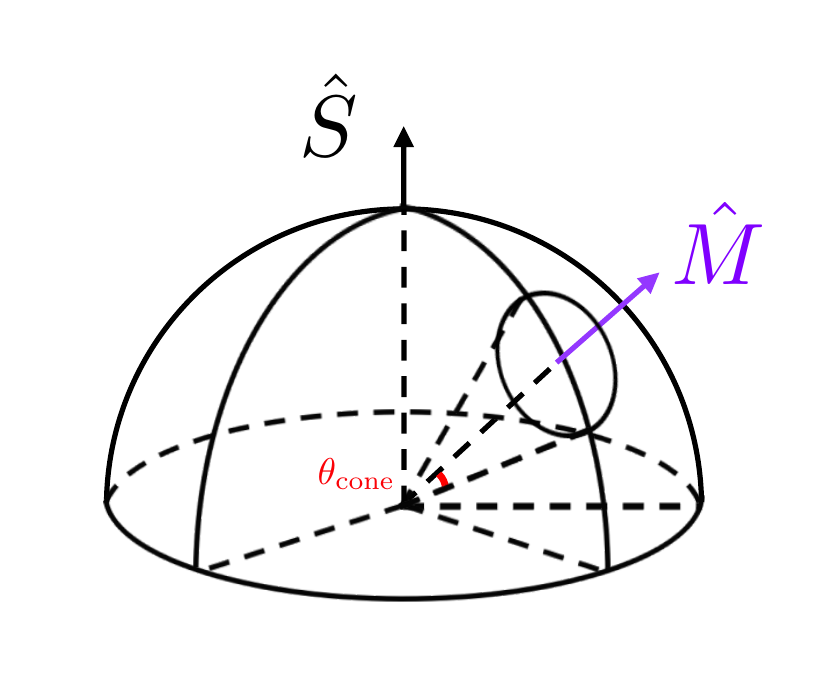}
        %\vspace*{1.05cm}          
    \caption{
    Geometry of the \ac{msp}'s emission. 
    % spin and mag
    $\boldsymbol{\hat S}$ denotes the spin axis of the \ac{msp}, $\boldsymbol{\hat M}$ is the magnetic axis and is rotating around $\boldsymbol{\hat S}$. The circles on the sphere surrounding the magnetic axis denotes the upper and lower boundary of the radiation beam of \ac{msp}, integrated in time. 
    % radiation beam 
    % Observer
    The precession of the spin axis is achieved by moving the observer $\boldsymbol{\hat O}$ on the red circle in the $x'-y'$ plane, while fixing the direction of the spin's axis. Note that, the nutation of spin axis is not included and demonstrated in the figure, 
      and can be achieved by perturbing the observer's circular trajectory in $z'$ direction.
    % observation
    The observer receives the pulses when the unit vector is in between the two circles of radiation beam, as shown by Eq.~\ref{eq-VariationOfPulse}.
    % two observables....
    % inclinition
    The $x'-y'$ plane is inclined at an angle $\iota$ with respect to the original $x-y$ plane. 
    This angle can be transformed into the inclination angle of the \ac{msp} orbit using Eq.~\ref{eq-MotionOfObserver}.
    }
    \label{fig:MSP_Observer_geometry}
\end{figure}

Besides gravitational wave, the spin-orbit coupling effect can be observed in pulsar timing observation. 
The correction to orbital precession would introduce extra shift of pulses received by a distant observer, and the spin precession and nutation could lead to the variation of pulse profiles, the detection of which has been shown to be possible \citep{Kerr2015}. 
To estimate the effect of spin's precession and nutation, we adopt a toy model, using the light house model of pulsar here, as shown in Fig.~\ref{fig:MSP_Observer_geometry}.
In Fig.~\ref{fig:MSP_Observer_geometry}, 
  we do not use the Euler angles defined above. 
Instead, we fix the spin vector, while moving the observer relative to the centre of \ac{msp}, 
  in a way that could mimic both precession and nutation. 
The spin axis $\boldsymbol{\hat S}$ is fixed in the $x-z$ plane, 
  such that the magnetic field line $\boldsymbol{\hat M}$ 
  is initially aligned with the $x$-axis, and rotate around the spin axis with period $P_{\rm s}  = 1\;\!{\rm ms}$. 
The angle between $\boldsymbol{\hat M}$ and $\boldsymbol{\hat S}$ is $\chi$.
When the spin axis precesses around the angular momentum $\boldsymbol{\hat L} = -\boldsymbol{\hat z'}$, 
  we move the observer on the $x'-y'$ plane (around $z'$) to mimic the precession. 
When the spin axis nutates slightly in $-\boldsymbol{\hat L} = \boldsymbol{\hat z'}$ direction, 
  we move the observer in $-z'$ direction to mimic nutation. 
Here the $x'-y'$ plane is inclined at an angle $\iota$ with respect to the original $x-y$. 
All the vectors used here, 
  including $\boldsymbol{\hat S}$, 
  $\boldsymbol{\hat M}$ and $\boldsymbol{\hat O}$ are unit vectors.

In order to simplify our model, 
  we use a non-rotating massive black hole in the following discussion. 
Suppose that the precession has an angular frequency $\Omega$, 
  from the results in Eq.~\ref{eq-nutationfrequency}, 
  the dominate angular frequency of nutation is also $2(\omega - \Omega)$, 
  and the dominate correction to nutation is of frequency $\Omega$, 
  as shown in Fig.~\ref{fig:SpinPrecessionTheta} and Fig.~\ref{fig:SpinPrecessionTheta2}. 
We write the nutation as $\nu(\Omega,t,\omega)$. 
The location of the observer at time $t$ is 
\begin{equation}
\begin{aligned} \label{eq-MotionOfObserver}
x_{\rm OB} =& \cos (t \Omega ) \cos (\nu (\Omega ,t,\omega )) \ ; \\
y_{\rm OB} =& \sin (\Omega t) \cos (\nu (\Omega ,t,\omega )) \cos \iota + \sin (\nu (\Omega ,t,\omega )) \sin \iota \ ;\\
z_{\rm OB} =& \sin (\Omega t) \cos (\nu (\Omega ,t,\omega )) \sin \iota - \sin( \nu (\Omega,t,\omega) ) \cos \iota \ .\\
\end{aligned}
\end{equation}
As the time scale of the precession is much larger than the rotation time scale, 
  we can first ignore the rotation, 
  and assume that whenever the unit vector of the observer (i.e. the line-of-sight) is inside the rings wrapping the magnetic field axis, 
  the observer would receive pulses, 
  with width equivalent to the arc length. 
The emission cone is assumed to have an half open angle $\theta_{\rm cone}$. 
The width of the pulse is therefore:
\begin{equation}
\begin{aligned} \label{eq-VariationOfPulse}
w = & 2 \sin \theta_{\rm cone}  \sqrt{1- \frac{\tan^2 \theta_{MO} }{ \tan^2 \theta_{\rm cone}  }}  \text{ when $\theta_{MO} \leq \theta_{\rm cone} $} \ ,\\
\end{aligned}
\end{equation}
where $\theta_{MO}$ is the angle between unit vectors $\boldsymbol{\hat O}$ and $\boldsymbol{\hat M}$, where $\boldsymbol{\hat M}$ is rotated such that it's in the same plane as $\boldsymbol{\hat S}$ and $\boldsymbol{\hat O}$. The angle  $\theta_{MO}$ can therefore be determined by
\begin{equation}
\begin{aligned}
\theta_{MO} = |\cos^{-1}(\boldsymbol{\hat S} \cdot \boldsymbol{\hat O}) - \chi| \ .
\end{aligned}
\end{equation}

When we include the the rotation of the \ac{msp}, 
  as shown by the upper panel of Fig.~\ref{fig:MSP_Observer_geometry}, 
  as the observer moves from $\boldsymbol{\hat O}$ to $\boldsymbol{\hat O'}$, 
  the emission it receives is triggered by magnetic field $\boldsymbol{\hat M'}$. 
Therefore, there is a shift of emission time, either delayed or advanced by 
\begin{equation}
\begin{aligned} \label{eq-TimeShift}
\Delta t  & = \frac{1}{\pi} \left|\; \sin^{-1} \left( \frac{\sin (\theta_{MM'}/2) }{\sin \chi} 
  \right) \; \right| P_{\rm s} \ , 
\end{aligned}
\end{equation}
where $\theta_{MM'}$ is the angle between $\boldsymbol{\hat M}$ and $\boldsymbol{\hat M'}$.
The angle $\theta_{MM'}$ can be calculated by using 
\begin{equation}
\begin{aligned}
\sin \left(\frac{\theta_{MM'}}{2}\right) & = \sqrt{\frac{1 -\cos \theta_{MM'} }{2}} = \sqrt{\frac{1 -\boldsymbol{\hat M} \cdot \boldsymbol{\hat M'} }{2}} \ ; \\ 
\boldsymbol{\hat M'} & =  \frac{\sin (\theta_{w} + \chi)}{\sin \theta_{w}}  \boldsymbol{\hat S} + \frac{\sin \chi}{\sin \theta_{SO}} (\boldsymbol{\hat O} - \boldsymbol{\hat S}) \ ,
\end{aligned}
\end{equation}
where $\theta_{w} = \left(\pi - \theta_{SO}\right)/2$, and $\theta_{SO} = \cos^{-1} (\boldsymbol{\hat S} \cdot \boldsymbol{\hat O})$.

\begin{figure}
	  \vspace*{0.05cm}   \center 
     \includegraphics[width=1\columnwidth]{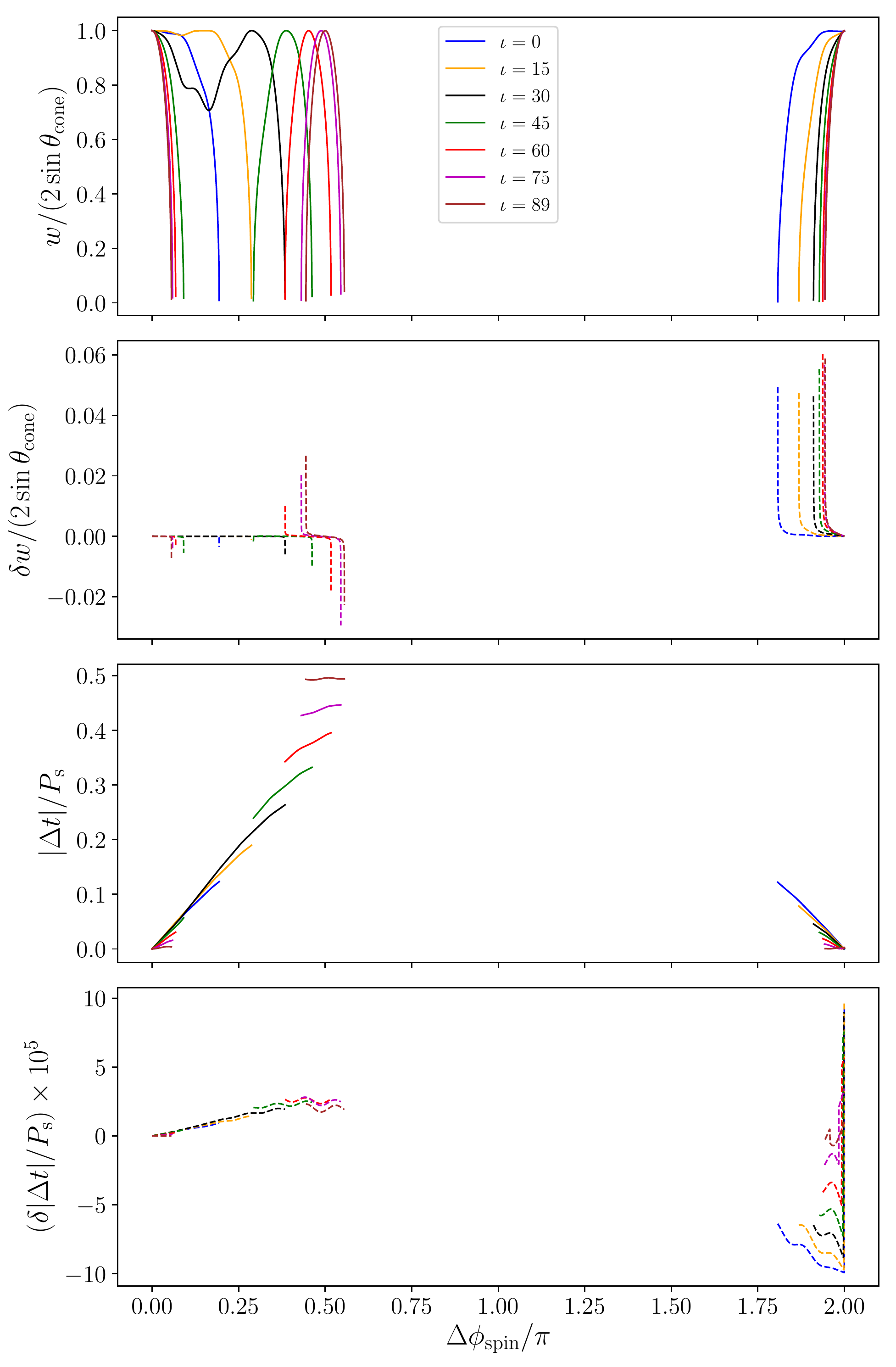}
        \vspace*{-0.25cm}          
    \caption{
    The $1$st and $3$rd panels show the relative width of each pulse $w$ (compared with the maximum possible width $2 \sin \theta_{\rm cone}$) and the shift of emission time $\Delta t$ (compared with \ac{msp}'s intrinsic spinning period $P_{\rm s}$ ) received by a distant observer moving inside $x'-y'$ plane.  
    The plane is inclined at different angle $\iota$ with respect to the $x-y$ plane. $\iota= 90^\circ $ was replaced by $89^\circ $ to avoid the coordinate singularity.  
    Only geodesics precession and nutation are included in the 
    The 2nd upper panel is the variation in width when including the effect of \ac{msp}'s spin, the dashed lines corresponds to the solid lines with the same colour in the 1st panel. The 4th panel is the corresponding corrections to the shift of emission time in 3rd panel due to \ac{msp}'s spin. 
    The corrections due to \ac{msp}'s spin is shown in the Table.~\ref{table:PulseEmissionShift}
}
    \label{fig:PulsesWidthAndTimeShift}
\end{figure}

Take $\theta_{\rm cone} = 10^{\circ}$, and $\chi = 45^{\circ}$ for an example. 
We used the data $(r=20M, a=0)$ 
from Fig.~\ref{fig:SpinPrecessionTheta2}. 
We consider only the leading order precession and nutation due to geodesics, 
  and the leading order correction due to the coupling of the \ac{msp}'s spin. 
We write the nutation as \begin{equation}
\begin{aligned} 
\nu (\Omega,t,\omega) = \theta_{\rm geo}  \left[ 1 - \cos 2 (\omega - \Omega )t \right] + 
\theta_{\rm cor} \left[ \cos  (\Omega t) - 1 \right] \ ,
\end{aligned}
\end{equation}
  where $\theta_{\rm geo}$ is the scale of nutation for geodesics motion, 
  $\theta_{\rm cor}$ is the scale of nutation, 
  and both are positive number. 
The values of $\omega, \Omega, \theta_{\rm geo}$ and $\theta_{\rm cor}$ for geodesics motion and the leading order corrections 
  are given in Table.~\ref{table:ValuesTable}. 
The variation in pulse width, and time shift are shown in Fig.~\ref{fig:PulsesWidthAndTimeShift}.

%%%%%%%%%%%%%%%%%%%%%%%%%%%%%%%%%%%%%%%%%%%%%%%%%%
% Table 2
\begin{center}
\begin{table}
\begin{tabular}{ |p{0.5cm}|p{3cm}|p{0.5cm}|p{3cm}} 
% \hline 
% \multicolumn{5}{|c|}{Prefactor of corrections due to spin-orbit coupling \label{table:TimeShift}} \\
 \hline
 & geodesics & & corrections \\
\hline
\hline
$\omega$ & $2.2739 ~ {\rm rad}/{\rm sec}$ & $\delta \omega$ & $-7.4411 \times 10^{-5} ~  {\rm rad}/{\rm sec} $ \\
\hline
$\Omega$ &  $0.17747 ~ {\rm rad}/{\rm sec}$ & $\delta \Omega$  &  $1.2599 \times 10^{-5} ~ {\rm rad}/{\rm sec}$ \\
\hline
$\theta_{\rm geo}$ & $7.1780 \times 10^{-3} ~ {\rm rad} $ & $\delta \theta_{\rm geo}$  & $0$ \\
\hline
$\theta_{\rm cor}$ & $0$ & $\delta \theta_{\rm cor}$ & $8.5512 \times 10^{-5} ~ {\rm rad} $ \\
\hline
\end{tabular}
 \caption{
The values of recession speed and nutation scale for data 
  $(r=20M, a=0, \theta_{\rm spin} = \pi/4, \phi_{\rm spin} = \pi)$ in Fig.~\ref{fig:SpinPrecessionTheta2}. 
Only leading order precession and nutation are considered in order to demonstrate the effect of \ac{msp}'s spin on the width and time shift of pulses. 
The corresponding results are shown in Fig.~\ref{fig:PulsesWidthAndTimeShift}. 
} \label{table:ValuesTable}
 \end{table}
\end{center}
%%%%%%%%%%%%%%%%%%%%%%%%%%%%%%%%%%%%%%%%%%%%%%%%%%

Besides changing the width and time shift of the pulse, 
the coupling of spin also results in the shift of pulse disappearing and appearing times, 
  i.e. the times when $w=0$ in the upper panel of Fig.~\ref{fig:PulsesWidthAndTimeShift}. 
The disappearing and appearing time, 
  denoted by $t_i$ are the $i$th solutions to $ \tan^2 \theta_{MO} =  \tan^2 \theta_{\rm cone}$, 
  which is equivalent to:
\begin{equation}
\begin{aligned}  \label{eq-SOequalChi}
f(t,\Omega,\omega,\theta_{\rm geo},\theta_{\rm cor}) \equiv \boldsymbol{\hat S} \cdot \boldsymbol{\hat O} = \cos (\chi \pm \theta_{\rm cone} )  \ ,
\end{aligned}
\end{equation}
   where we adopt the geodesics values of $(\Omega,\omega,\theta_{\rm geo},\theta_{\rm cor})$ as in Table.~\ref{table:ValuesTable}. 
The variations of $t_i$ can be found by solving the equation Eq.~\ref{eq-SOequalChi} 
   with $\omega \rightarrow \omega + \delta \omega $, $\Omega \rightarrow \Omega + \delta \Omega $, $\theta_{\rm geo} \rightarrow \theta_{\rm geo} +  \delta \theta_{\rm geo}$ and $\theta_{\rm cor} \rightarrow  \theta_{\rm cor} + \delta \theta_{\rm cor}$. 
The variations $\delta t_i$ are shown in Table.~\ref{table:PulseEmissionShift}. 
Such a shift is possible to be detected with the instrumental precision of pulsar timing \footnote{Currently 
  the upper limit of the pulsar timing precision of The Square Kilometer Array (SKA) is expected to be about 
  $\sim 10 - 100 \;\!{\rm ns}$ \citep[see e.g.][]{Stappers2018} and 
  the Five hundred meter Apeture Spherical Telescope (FAST) 
 $\sim 100\;\!{\rm ns}$, over 10-min time integration \citep[see e.g.][]{Hobbs2014}.   
Note that the precision is limited by the timing technique
  and it will improve accordingly with the further advancements of timing techniques and system modelling  
  \citep[see][]{Hobbs2006,Oslowski2011,Hobbs2014}.   
} that we can achieve in the near future. 
   
The time shift in Table.~\ref{table:PulseEmissionShift} is valid for MBH between $10^3 - 10^6 {\rm M}_{\odot}$, 
% results from SolveTimeNumericallyWithNutation2nd.py and SolveTimeNumericallyWithNutation.py
 and accumulates with about $\sim - 2.5 \mu {\rm s}$ every spin's precession period $T = {2\pi}/{\Omega}$, regardless of the mass of massive black hole. 

%%%%%%%%%%%%%%%%%%%%%%%%%%%%%%%%%%%%%%%%%%%%%%%%%%
% Table 3
\begin{center}
\begin{table}
\begin{tabular}{ |p{0.5cm}||p{1.5cm}|p{1.5cm}|p{1.5cm}|p{1.5cm}|} 
 \hline
$\iota$ & $\delta t_1 ({\rm ns}) $ & $\delta t_2 ({\rm ns}) $ & $\delta t_3 ({\rm ns}) $ & $\delta t_4 ({\rm ns}) $ \\
\hline
\hline
0  & -38.72 &   &   & -3613.66 \\
\hline
15 & -32.88 &   &   & -2887.72 \\
\hline
30 & -141.64 &   &  & -2113.76 \\
\hline
45 & -185.35 & 23.34 & -224.91 & -1987.59 \\
\hline
60 & -77.88 & -242.83 & -368.60 & -2246.72 \\
\hline
75 & -70.66 & -391.80 & -539.21 & -2397.76 \\
\hline
89 & -69.65 & -541.13 & -685.78 & -2441.96 \\
\hline
\end{tabular}
 \caption{
 % results from Paper/SolveTimeNumericallyWithNutation.py
 The variation in the disappearing and appearing time of the pulses. 
Here $t_{1,2,3,4}$ correspond to the instants, when the pulse signal 
  (1) first leaves the emission cone for the first time,
  (2) re-enter the emission cone for the first time,
  (3) re-leave th emission cone for the second time,
  (4) and re-enter the emission cone for the second time within one precession period (i.e. within $0 < \Omega t < 2 \pi$). 
Note that for $\iota=0,15,30$, 
  the observer's line of sight doesn't re-enter the emission cone until the end of a precession period, 
  therefore, we will skip $t_2$, $t_3$, 
  and label the first re-entering time as $t_4$. 
 } \label{table:PulseEmissionShift}
 \end{table}
\end{center}
%%%%%%%%%%%%%%%%%%%%%%%%%%%%%%%%%%%%%%%%%%%%%%%%%%

%%%%%%%%%%%%%%%%%%%%%%%%%%%%%%%%%%%%%%%%%%%%%%%%%%
% Figure 7
%%%%%%%%%%%%%%%%%%%%%%%%%%%%%%%%%%%%%%%%%%%%%%%%%%

%%%%%%%%%%%%%%%%%%%%%%%%%%%%%%%%%%%%%%%%%%%%%%%%%%
% Figure 8
%%%%%%%%%%%%%%%%%%%%%%%%%%%%%%%%%%%%%%%%%%%%%%%%%%

%%%%%%%%%%%%%%%%%%%%%%%%%%%%%%%%%%%%%%%%%%%%%%%%%%
\subsection{Pulsars orbiting around a massive black hole} 
\label{sec:Events}
{ Large population of pulsars are believed to reside in the central region of our galaxy \citep{Pfahl2004,Wharton2012,Zhang2014},
  and pulsar population is believed to be dominated by \acp{msp}, 
  the species that existing pulsar searches are not sensitive to \citep{Macquart2015}.
Several indirect pieces of evidence have been supporting this prediction,
  for example, excessive gamma-ray emission \citep{Brandt2015}
  (although dark matter could as well account for such gamma-ray detections),
  the detection of rare magnetar \citep{Rea2013,Eatough2013,Mori2013}, 
  and the dense stellar environment in the galactic centre.   
Very few pulsars have been found until now, and it is believed that the strong scattering of the radio wave by interstellar medium and severe dispersion
  along the line-of-sight reduce the chance for pulsars to be detected \citep{Cordes1997,Lazio1998}.
Especially, for \acp{msp}, the temporal smearing at low frequencies is severe \citep{Macquart2010}, 
  and current detectors are not sensitive enough to detect \acp{msp} as they are of low luminosity at high frequency \citep{Bower2018}. 
Despite the null detection, prediction of the pulsar population in the central region of our galaxy has
  been made with different models and assumptions.
Constraints from gamma-ray and radio observations predicted $\leq 10^3$ \ac{msp} inside $1 \, {\rm pc}$ \citep{Wharton2012},
  and up to $10^4$ \acp{msp} inside $1 \, {\rm pc}$ \citep[depending on the scattering and absorption, see][for details]{Rajwade2017}.
Simulations by \citet{Zhang2014} predicted $\sim 10$ pulsars inside $\leq 1000 \, {\rm au}$, where they assumed that massive 
  stars were captured by central black hole by tidal disruption of stellar binaries.   
%It's expected that high sensitivity detectors, 
%  e.g. \ac{ska} in the future are expected to open up a new window for detecting \acp{msp} 
%  in the galactic centre \citep{Pfahl2004,Torne2017}. 
However, these studies on pulsar in the galactic centre are focused on the non-relativistic regime, 
  where the effect of the pulsar's spin is not important. 
Indeed, the event rate of discovering a pulsar in the close vicinity of our galactic nuclear black hole is quite low. The complicated environment in 
  the galactic centre makes pulsars difficult to be detected, even if they exist.

Pulsars on orbits with an intermediate-mass black hole (i.e. \acp{imrb}) are potentially
  more promising sources. }
Observations have shown that large galaxies contain a massive nuclear black hole 
   and some may have two, e.g.\ M83 \citep{Thatte2000}.   
The masses of these nuclear black holes are found to correlate 
  with the dynamical properties, and hence the mass, of spheroid components of the host galaxies 
   \citep{Magorrian1998,Ferrarese2000,Gebhardt2000}. 
Although the empirical correlations may deviate 
   at the low-mass end where the small galaxies are located 
  \citep[see][]{Graham2015}, 
  it does not exclude that 
  ultra-compact dwarf galaxies, globular clusters, or million solar-mass stellar spheroids 
  would contain a black hole at the centre \citep[see]{Perera2017}. 
The masses of the black holes residing in these spheroids 
  are expected to be $\sim 10^2 - 10^4~{\rm M}_\odot$ \citep{Lutzgendorf2013,Mieske2013}, 
  distinguishing them from the stellar-mass black holes in X-ray binaries,  
  e.g. GRO~J1655$-$40 \citep[see][]{Soria1998,Shahbaz1999} 
  LMC X-3 \citep[see][]{Orosz2014} in the nearby universe.   
It is still unclear how and whether a massive nuclear black hole would be formed   
  at the central region of ultra-compact dwarf galaxies or globular clusters.  
A black hole can grow by accreting gas or capturing stars. 
In dense stellar environments such as the central region of a ultra-compact dwarf galaxy  
  or the core of a globular cluster, a nuclear black hole, if it is present, could gain mass 
  by coalescing with another black hole, if it is also present, or 
  by capturing stars in its neighbourhood.  
The recent \ac{ligo} observations have confirmed 
  that a more massive black hole can be formed by the coalescence of two smaller-mass black holes 
  \citep[e.g.][]{Abbott2016,Abbott2017b} 
  and a black hole can be produced by merging two neutron stars \citep{Abbott2017a}. 
Naturally, we can generalise that 
  a more massive black hole can also be formed 
  by merging with a neutron star or a black hole, 
  though such events have not been observed yet. 
{ Some studies \citep[e.g.][]{Fragione2018} indicated
  that merging of two black holes in a globular cluster would likely cause the remnant system to be ejected.
It is, therefore, a concern
  whether a nuclear black hole would grow to $10^3~{\rm M}_\odot$ through a
  sequence of  
  black-hole black-hole merging process.
Observational studies, however, have shown support for 
  the presence of intermediate-mass black holes \citep[see e.g.][]{Feng2011} in a number of external galaxies.
There were also claims \citep[][]{Perera2017}   
  that intermediate-mass black holes were found in globular clusters. 
However, these pieces of evidence are not conclusive. 
Whether or not globular clusters can retain a nuclear intermediate-mass black hole 
   and hence a location of \acp{imri} would be better resolved by future 
   multimesenger studies, 
   using instruments such as SKA \citep[see][]{Wrobel2018} and \ac{lisa} \citep[see e.g.][]{Kimpson2019b,Kimpson2019a}.  }

\acp{msp} are fast spinning neutron stars on a period  
  of a few to about ten~ms 
  \citep[see e.g.][]{Bhattacharya1991,Papitto2014} 
Some newly born neutron stars have a spin period of as short as a few tens of milliseconds, 
  e.g.\ the Crab pulsar with a spin period of 33~ms \citep[see][]{Manchester2005}.   
\acp{msp} are however old neutron stars, found in globular clusters and the galactic bulges. 
They are believed to have a binary progenitor, 
   and the neutron star was spun up through accreting matter from a companions star 
   \citep{Radhakrishnan1982,Bhattacharya1991,Ergma1996,Tauris2000}. 
Many globular clusters are rich in \ac{msp} --  
 25 have been identified in 47~Tucanae \citep[see][]{Manchester1991,Pan2016} 
  and 33 in Terzan~5 \citep[see][]{Ransom2005,Hui2010}. 
With the abundant \ac{msp} populations,   
  { \ac{imrb}} comprising an \ac{msp} and a black hole could be formed in the core of a globular cluster    
  \citep{Devecchi2007,Clausen2014,Verbunt2014}. 
An estimate of $\sim 1-10$ these \ac{msp} - black hole binaries 
  in the Galactic globular clusters \citep[see][]{Clausen2014} 
  would imply that a few tens of such binaries could reside  
  in the globular clusters in the Local Group galaxies, 
  and the pulse emission from the \ac{msp} could be detected 
  by large ground-base radio telescopes 
  such as the The Square Kilometer Array (SKA) \citep{Keane2015} and Five hundred meter Apeture Spherical Telescope 
(FAST) \citep{Nan2006}.  

Although the gravitational radiative loss 
  would have insignificantly effects on the spin and orbital dynamics of the \ac{msp} 
  in the \ac{emrb}{ /\ac{imrb}} considered in this work \cite[see][]{Singh2014},  
  the power of the gravitational waves emitted from these systems is not negligible. 
For a system with a black hole with  
   a mass $M = 10^3\;\!{\rm M}_\odot$, a spin parameter $a = 0$, 
   and an \ac{msp} - black hole orbital separation $r =  20\;\!M$,  
   the gravitational wave power could reach $\sim 1.6\times 10^{48}\;\!{\rm erg~s}^{-1}$ 
   assuming a circular orbit. 
The corresponding gravitational wave strain $h$ is $3.5 \times 10^{-18}$, 
   if the system is located at the core of a globular cluster 
   at a distance of 5~kpc from the Sun, similar to that of 47~Tucanae \citep{Carretta2000}.  
These systems, which are persistent gravitational wave sources, 
  will eventually evolve to become \ac{emri}{ /\ac{imrb}} burst gravitational wave sources,  
  when the \ac{msp} spirals in and coalesces with the black hole.  
They are expected to be detectable 
  within the \ac{lisa} band in the \ac{emri}{ /\ac{imri}} stage and also in the \ac{emrb}{ /\ac{imrb}} stage.    

The significance of these \ac{emrb}{ /\ac{imrb}} sources 
  in the context of gravitational wave and multi-messenger astrophysics are of two folds. 
First of all, 
  the statistics of the \ac{emri}{ /\ac{imri}} events arisen from these systems 
  and of the detection of them in the \ac{emrb}{ /\ac{imrb}} { phase} 
  will provide us a mean to determine the abundances of these systems 
  and their populations in various galactic environment. 
This in turn will constrain their formation channels 
  in dense stellar systems with a resident black hole. 
Secondly,  
  knowing the population of \ac{msp} - black holes binaries in globular clusters or other dense stellar spheroids 
  would provide an estimate 
  of the number detectable individual persistent gravitational wave sources,   
  and hence their contribution to the stochastic gravitational wave background. 
It will serve as a reference when we build models to compute the \ac{emri}{ /\ac{imri}} events 
  arisen from neutron star - black hole binaries 
  in the less understood dense stellar environment in the distant Universe.

%%%%%%%%%%%%%%%%%%%%%%%%%%%%%%%%%%%%%%%%%%%%%%%%%%
\subsection{Additional remarks} 

In this work, we assume that
  the \ac{msp} is a point pole-dipole, moving in the static Kerr space time. 
However, in realistic situation, the \ac{msp} will also 
  curve the space time around it and the background
  space time will be the consequence of non-linear 
  combination of the \ac{msp}'s gravity with the black hole's gravity. 
The trajectory of the \ac{msp} will be the geodesics (if we ignore spin-orbit and spin-curvature couplings)
  of such complicated and evolving space time. 
Therefore, it's necessary for us to verify the effect of the \ac{msp}'s own gravity (so called self-force)
  on the orbital dynamics and spin dynamics.  

{
The investigation into the effects of the self-force 
 and its comparison with the spin-orbit coupling force
  has been carried out extensively \citep[e.g.][]{Burko2004,Bini2014,Bini2015,Burko2015,Barack2018}
  in different contexts.
The magnitude of the first order self-force (in terms of mass ratio $m/M$) is similar to that of the
  spin-orbit couplings \citep{vandeMeent2018}
  \citep[also as shown by comparing results in][with Fig.~\ref{fig:Periastron}]{Barack2011}. 
The leading term of the correction to the rate of the spin's precession due to the conservative part of
  the first order self-force is \citep[adapted from Eq.~(10) or equivalently Eq.~(5.4) of][respectively]{Dolan2014,Bini2014}  
}
\begin{equation}
\begin{aligned}  
\omega_{1{\rm st}} \propto \frac{m}{M} \frac{M}{r} \omega_{L_{\rm N}} \ ,
\end{aligned}
\end{equation}
    which is smaller than that due to \ac{msp}'s spin 
    (i.e. Eq.~\ref{eq:approximatespinprecession}) by a factor of ${M}/{r}$. 
This self-force correction could be important 
  if we integrate the pulsed signal over a substantial duration.  

% order of GW emission

The leading order of the dissipative self-force is also called radiation-reaction,
  and it introduces the loss of energy and angular momentum in the \ac{emri}{ /\ac{imri}} system \citep{Barack2018}. 
{ The energy flux and angular momentum flux have been calculated by \citep{Drasco2006,Fujita2009,Fujita2012,Shah2014,vandeMeent2018}
  for different orbital configurations.  
Contribution to the dephasing of \ac{gw} waveform from dissipative
  self-force is in general greater than that from conservative self-force
  and spin-orbit couplings \citep{Burko2015}. 
To the lowest order, the dissipative self-force can be calculated by solving the 
  energy and angular momentum balance equations \citep{Barack2007,Burko2013}, 
  and we could estimate its effects using the radiative loss formula in \cite{Peters1964}}
{ \footnote{It worth noticing that, the energy loss rate calculated by \cite{Peters1964}
  is based on the assumption that the binary follows a non-precessing Newtonian eccentric orbit,
  and energy flux is integrated over an infinite distant sphere enclosing the binary.   
  However, for \ac{emri} systems, the energy is calculated by solving black hole's perturbation equation 
  and energy flux into the black hole's horizon is also considered 
  (which is substantially smaller than the energy flux to infinity \citep{Barack2007}).     
These two schemes are equivalent only to the Newtonian order and lowest order of mass ratio (i.e. $\propto (m/M)^2$).  
}}.

From the spin precession angular frequency in Eq.~\ref{eq:spinprecessionnewtonian}, 
  we may define a spin precession period:   
\begin{equation} 
\begin{aligned} 
P_{\rm sp} \approx \frac{2 \pi}{ \langle \omega_{L_{\rm N}} \rangle} = \frac{ 4 \pi}{3} \frac{r^{5/2} }{M^{2}} \sqrt{m + M} \left(1 - e^2\right) \ ,
\end{aligned}
\end{equation} 
where $\langle \omega_{L_{\rm N}} \rangle$ denotes the average value of $\omega_{L_{\rm N}}$ over an orbital period, 
  under the approximation that the \ac{msp} follows a Newtonian eccentric orbit. 
The time-scale for the change in the spin's precession period due to gravitational 
  radiation is 
\begin{equation} 
\begin{aligned} 
\tau_{\rm gw} \sim \left[ \frac{ r^4}{32 m M (m +M)}\right]\, g(e)^{-1}   
\end{aligned}
\end{equation} 
\citep[see][]{Peters1964}, 
where $r$ is the semi-major axis, and $g(e)$ is a function of orbital eccentricity:
\begin{equation} 
\begin{aligned} 
g(e) = (1-e^2)^{-7/2} \left( 1+\frac{71}{40} e^2-\frac{19}{160} e^4  \right)  \ . 
\end{aligned}
\end{equation} 
Setting $r = \zeta M$, we have
\begin{equation} 
\begin{aligned} 
\frac{ \tau_{\rm gw}}{P_{\rm sp}} \sim \frac{3}{128\pi} \frac{\zeta^{3/2} }{g(e) \left(1-e^2\right)} \left(\frac{M}{m}\right) \left(\frac{M}{m+M}\right)^{3/2} \ .  
\end{aligned}
\end{equation} 
For the systems considered here,     
  $10^3 {\rm M}_\odot \le M \le 10^5\;\! {\rm M}_\odot$ and 
  $20 \le \zeta \le 100$,  
  implying that ${M}/{m} > 6 \times 10^2$.      
Moreover, the orbital eccentricity $e \le 0.4$.  
This gives $g(e) (1-e^2) \sim (1-2)$, 
  and ${\tau_{\rm gw}}/{P_{\rm sp}} \sim (2\times 10^2 - 5 \times 10^5) \gg 1 $.
The application of the \ac{mpd} formulation is therefore justified.   

{ We would like to emphasize that, the formula above are only valid
  for orbits with moderate eccentricities.    
For highly eccentric orbits, ignoring the effects of self-force would lead to 
  substantial errors in modelling the orbital dynamics of the \ac{msp}.   
In both cases,}
  including the effects of the self-force will be necessary
  for modelling secular evolution and the orbital dynamics of \ac{msp} in an \ac{emrb}{ / \ac{imrb}} with high temporal resolutions.

% modeling difficulties
Besides the time shift and width variation of the pulses 
  due to the precession and nutation of the spin,  
  as that shown in Sec.~\ref{sec:observation}, 
  and the orbital deviation from geodesic motion \citep[studied in][]{Singh2014}, 
  the bending of light (i.e. gravitational lensing) due to the black hole's gravity can be non-negligible. 
To achieve the scientific goals described in this work, 
  we need high temporal and spatial accuracies in the covariant photon transport calculations. 
A self-consistent calculation as such is computationally challenging and  
  and it also requires advanced numerical techniques, 
  and hence it is beyond the scope of the semi-analytic approached adopted this work.  
We leave such calculations to future studies.

%%%%%%%%%%%%%%%%%%%%%%%%%%%%%%%%%%%%%%%%%%%%%%%%%%
%%%%%%%%%%%%%%%%%%%%%%%%%%%%%%%%%%%%%%%%%%%%%%%%%%
\section{Summary and conclusion}
\label{sec:summary}

We investigate the spin dynamics of an \ac{msp} around a massive black hole using the 
  \ac{mpd} formulation. 
The extreme mass ratio of the system allows us to consider that 
  the \ac{msp} is a spinning test mass in a spacetime provided by the black hole.  
The orbital motion can be described as quasi-geodesics 
  with corrections due to spin-orbit, spin-spin and spin-curvature couplings. 
These spin couplings lead to precession and nutation of the \ac{msp}'s spin, 
  besides perturbing the \ac{msp}'s orbital motion. 
Such modulations will be detectable in the future gravitational wave experiments, 
  such as \ac{lisa}, 
  and in pulsar timing observations, with instruments such as SKA and FAST. 
We have also shown that, the spin-orbit and spin-spin couplings 
  will lead to timing variations 
  between the reference frame of the \ac{msp} and the observer at a long distance.  
The timing variation will manifest as variations in the pulsed periods 
  of the pulsar's emission received by the observer. 
These results obtained from \ac{mpd} equations are consistent 
  in order with the weak field approximation.

%%%%%%%%%%%%%%%%%%%%%%%%%%%%%%%%%%%%%%%%%%%%%%%%%%
%%%%%%%%%%%%%%%%%%%%%%%%%%%%%%%%%%%%%%%%%%%%%%%%%%
\section*{Acknowledgements}
We thank P.~K.~Leung for insightful discussions 
  on relativistic dynamics,  
  J.-L.~Han on pulsar observations, 
  and A.~Gopakumar and T.~G.~F.~Li on orbital eccentricities in extreme-mass-ratio binary systems.    
We also thank Patrick C.~K.~Cheong and P.~K.~Leung 
  for suggestions, comments and critical reading of the manuscript. 
KW thanks the hospitality of National Astronomical Observatory, Chinese Academy of Sciences 
   and CUHK Department of Physics,   
   where part of this work was carried out, during his visits.  
KJL's research at UCL MSSL was supported 
  by UCL through a MAPS Dean's Research Studentship   
  and by CUHK through a C.~N.~Yang Scholarship, 
  a New Asia College Scholarship,  
  a University Exchange Studentship, 
  a Science Faculty Research Studentship { and
  a Physics Department SURE Studentship.}

%%%%%%%%%%%%%%%%%%%%%%%%%%%%%%%%%%%%%%%%%%%%%%%%%%



\bibliographystyle{mnras}
\bibliography{reference}

%%%%%%%%%%%%%%%%%%%%%%%%%%%%%%%%%%%%%%%%%%%%%%%%%%
%%%%%%%%%%%%%%%%%%%%%%%%%%%%%%%%%%%%%%%%%%%%%%%%%%

%%%%%%%%%%%%%%%%%%%%%%%%%%%%%%%%%%%%%%%%%%%%%%%%%%

%%%%%%%%%%%%%%%%% APPENDICES %%%%%%%%%%%%%%%%%%%%%

\appendix

\section{Eccentricity} \label{EccentricityCalculation}

The eccentricity of the \ac{msp}'s orbit is set by solving the set of equations:
\begin{equation}
\begin{aligned}
E^2 {g_{\rm a}}^{00}-2 {g_{\rm a}}^{30}  E  L_z+{g_{\rm a}}^{33} L_z^2  =&  -1 \\
E^2 {g_{\rm b}}^{00}-2 {g_{\rm b}}^{30}  E  L_z+{g_{\rm b}}^{33} L_z^2  =&  -1 \ ,
\end{aligned}
\end{equation}
with ${g_{\rm a}}^{\mu\nu}$ evaluated at $r_{\rm a} = r(1-e)$ and ${g_{\rm b}}^{\mu\nu}$ at $r_{\rm b} = r(1+e)$, 
  where $r$ is the semi-major axis, and $e$ is the orbit eccentricity.  

The expressions for the solutions are complicated. 
We therefore include only the leading order of $\eta = {M}/{r}$ and $e$, 
  so as to demonstrate the leading order effect the of eccentricity and black hole's spin. 
We use $\alpha = {a}/{M}$, which is a dimensionless factor of the black hole spin. 
For prograde motion, the two integration constants, $E$ and $L_{z}$, associated with the geodesics are:
\begin{equation}
\begin{aligned} \label{eq-progradeEAM}
E & = 1-\frac{\eta }{2}+\frac{3 \eta ^2}{8}-\alpha  \eta ^{5/2} + \mathcal{O} \left( \eta^{5/2}, e^2 \right) \ , \\
L_z & = \frac{M}{ \sqrt{\eta }}  + \mathcal{O} \left( \eta^{-1/2}, e^2 \right)  \ , 
\end{aligned}
\end{equation}
and for retrograde motion, 
\begin{equation}
\begin{aligned} \label{eq-retrogradeEAM}
E & = 1-\frac{\eta }{2}+\frac{3 \eta ^2}{8}+5 \alpha  \eta ^{5/2}+12 e \alpha  \eta ^{5/2} +  \mathcal{O} \left( \eta^{5/2}, e^2 \right)  \ , \\
L_z & = \frac{M}{ \sqrt{\eta }}  + \mathcal{O} \left( \eta^{-1/2}, e^2 \right)  \ .
\end{aligned}
\end{equation} 

By setting the initial $E$ and $L_{z}$, the eccentricity is accurate for pure geodesic calculations 
 (i.e. setting $\lambda=0$ in Eq.~\ref{eq:mpd-momentum},\ref{eq:mpd-spin},\ref{eq:orb-coup}). 
 In such a situation, the 4-momentum and 4-velocity are parallel to each other. 
For the \ac{mpd} equation (Eq.~\ref{eq:mpd-momentum},\ref{eq:mpd-spin},\ref{eq:orb-coup} with $\lambda=1$), 
the eccentricity is slightly different from the expected values by $\delta e \sim 2 \times 10^{-5}$, 
which can be estimated by the comparison of the spin-orbit coupling force with the Newtonian gravitational force:
\begin{equation}
\begin{aligned} 
\frac{F_{\rm so}}{F_{\rm New}} \sim \left( \frac{M}{r} \right)^{3/2} \left( \frac{m}{M} \right) \frac{S}{m^2} \ .
\end{aligned}
\end{equation}

\section{Proper Time} \label{ProperTimeCalculation}

In order to investigate the effect of GR, eccentricity and black hole's spin on the ratio of the coordinate time over proper time  
${{\rm d}t}/{{\rm d}\tau}$, we calculate the approximate $u^0$ for quasi-circular orbits, using the $E$ and $L_z$ derived in  Appendix~\ref{EccentricityCalculation}. 

For prograde motion, the time component of the 4-velocity at $r_{1,2} = r(1 \mp e)$ is:
\begin{equation}
\begin{aligned} 
u^0 & = - g^{00} E - g^{03} E L_z \\
& = 1 + \left(\frac{3}{2} \pm 2 e\right) \eta  + \left(\frac{27}{8} \pm 7 e \right) \eta ^2  - \alpha \eta ^{5/2} (3 \pm 6 e ) + \mathcal{O}\left(\eta^3, e^2 \right) \ ,
\end{aligned}
\end{equation}
where the upper signs denote $r_1$, lower signs denote $r_2$. For retrograde motion, the $u^0$ is equivalent to changing $\alpha$ into $-\alpha$, as we can expected. 

This equation doesn't include the effect of \ac{msp}'s spin on the orbital $u^0$. Such effect can be estimated by means of the formula of spin-orbit coupling force as in Appendix~\ref{EccentricityCalculation}. As the eccentricity is shifted by $\delta e \sim 2 \times 10^{-5}$, the shift of $u^0$ is about 
\begin{equation}
\begin{aligned} 
\Delta u^0 \sim  2 \delta e \eta  \ .
\end{aligned}
\end{equation}

%%%%%%%%%%%%%%%%%%%%%%%%%%%%%%%%%%%%%%%%%%%%%%%%%%

% Don't change these lines
\bsp	% typesetting comment
\label{lastpage}
\end{document}